\documentclass[sigconf, nonacm]{acmart}

\newcommand\vldbdoi{10.14778/3503585.3503593}
\newcommand\vldbpages{XXX-XXX}
\newcommand\vldbvolume{15}
\newcommand\vldbissue{4}
\newcommand\vldbyear{2022}
\newcommand\vldbauthors{\authors}
\newcommand\vldbtitle{\shorttitle} 
\newcommand\vldbavailabilityurl{}
\newcommand\vldbpagestyle{empty} 

\setcopyright{none}
\renewcommand\footnotetextcopyrightpermission[1]{} 

\usepackage{stfloats}
\usepackage{multirow}
\usepackage{xcolor}
\newcommand{\feedback}[1]{\textcolor{black}{#1}}
\newcommand{\edit}[1]{\textcolor{black}{#1}}

\newcommand{\mv}[1]{\textcolor{black}{#1}}
\newcommand{\mvrev}[1]{\textcolor{black}{#1}}
\newcommand{\mvpolish}[1]{\textcolor{black}{#1}}
\newcommand{\dk}[1]{\textcolor{black}{#1}}
\newcommand{\dkpolish}[1]{\textcolor{black}{#1}}

\newcommand{\reals}{\mathrm{I\! R}}

\newcommand{\E}{\mathrm{I\! E}}
\newcommand{\EC}[2] {\mathrm{I\! E}\!\left[\left.#1\right|#2\right]}
\newcommand{\VC}[2] {\mathrm{Var}\!\left[\left.#1\right|#2\right]}
\newcommand{\PC}[2] {\mathrm{Pr}\!\left[\left.#1\right|#2\right]}

\begin{document}

\title{Popularity Prediction for Social Media\\ over Arbitrary Time Horizons}

\author[1]{Daniel Haimovich}
\affiliation{
	\institution{Core Data Science, Meta Platforms}
}
\email{danielha@fb.com}

\author[1]{Dima Karamshuk}
\affiliation{
	\institution{Core Data Science, Meta Platforms}
}
\email{karamshuk@fb.com}

\author[1]{Thomas J. Leeper}
\affiliation{
	\institution{Core Data Science, Meta Platforms}
}
\email{thomasleeper@fb.com}

\author[1]{Evgeniy Riabenko}
\affiliation{
	\institution{Core Data Science, Meta Platforms}
}
\email{riabenko.e@gmail.com}

\author[1]{Milan Vojnovic}
\affiliation{
	\institution{Department of Statistics, LSE}
}
\email{m.vojnovic@lse.ac.uk}

\begin{abstract} 
Predicting the popularity of social media content in real time requires approaches that efficiently operate at global scale. Popularity prediction is important for many applications, including detection of harmful viral content to enable timely content moderation. The prediction task is difficult because views result from interactions between \mvpolish{user interests}, content features, resharing, feed ranking, and network structure. We consider the problem of accurately predicting popularity both at any given prediction time since a content item's creation and for arbitrary time horizons into the future. In order to achieve high accuracy for different prediction time horizons, it is essential for models to use  static features (of content and user) as well as observed popularity growth up to prediction time. 

We propose a feature-based approach based on a self-excited Hawkes point process model, which involves prediction of the content's popularity at one or more reference horizons in tandem with a point predictor of an effective growth parameter that reflects the timescale of popularity growth. This results in a highly scalable method for popularity prediction over arbitrary prediction time horizons that also achieves a high degree of accuracy, compared to several leading baselines, on a dataset of public page content on Facebook over a two-month period, covering billions of content views and hundreds of thousands of distinct content items. The model has shown competitive prediction accuracy against a strong baseline that consists of separately trained models for specific prediction time horizons.
\end{abstract}

\maketitle

\pagestyle{\vldbpagestyle}
\begingroup\small\noindent\raggedright\textbf{PVLDB Reference Format:}\\
\vldbauthors. \vldbtitle. PVLDB, \vldbvolume(\vldbissue): \vldbpages, \vldbyear.\\
\href{https://doi.org/\vldbdoi}{doi:\vldbdoi}
\endgroup
\begingroup
\renewcommand\thefootnote{}\footnote{\noindent
This work is licensed under the Creative Commons BY-NC-ND 4.0 International License. Visit \url{https://creativecommons.org/licenses/by-nc-nd/4.0/} to view a copy of this license. For any use beyond those covered by this license, obtain permission by emailing \href{mailto:info@vldb.org}{info@vldb.org}. Copyright is held by the owner/author(s). Publication rights licensed to the VLDB Endowment. \\
\raggedright Proceedings of the VLDB Endowment, Vol. \vldbvolume, No. \vldbissue\ %
ISSN 2150-8097. \\
\href{https://doi.org/\vldbdoi}{doi:\vldbdoi} \\
}\addtocounter{footnote}{-1}\endgroup

\ifdefempty{\vldbavailabilityurl}{}{
\vspace{.3cm}
\begingroup\small\noindent\raggedright\textbf{PVLDB Artifact Availability:}\\
The source code, data, and/or other artifacts have been made available at \url{\vldbavailabilityurl}.
\endgroup
}

\section{Introduction}

Popularity prediction \feedback{can be a useful} system component for management of user-generated content in online platforms. For example, in content moderation platforms, such as the one used by Facebook \cite{FB20}, potentially harmful content items are flagged either by users or machine learning filters. These flagged content items are examined either automatically or are placed into a queue for manual review. To make sure that the most important posts are seen first by the reviewers, a content moderation platform may take into account their virality. Other applications of popularity prediction include optimizing content distribution, e.g. for video streaming \cite{Tang17}. In these applications, popularity prediction is used to prioritize content item processing with the goal to improve the quality of user experience. These applications require accurate and scalable methods for popularity prediction. 

State of the art popularity prediction algorithms are accurate but \edit{mostly} do not scale to handle \feedback{large-scale} social media content workload because they \edit{typically} have per-content-item computation cost that increases linearly with the number of observed events \edit{(see discussion in Sec.~\ref{sec:complexity})}. While different popularity prediction methods have been proposed, e.g., \cite{Chen13b,CADKL14,Zhao2015,HIP17}, they do not satisfy at least one of the following design \feedback{considerations} for application at a planetary scale: (a) prediction of the number of views acquired up to a future time horizon, not just a classification of virality, (b) prediction method has a low computation and memory complexity, (c) prediction method can generate accurate predictions for any given time horizon, or (d) prediction method leverages both static features (e.g. content author and content item features) and temporal features (observed up to given prediction time). We discuss this further as follows.

First, some work in the information cascades literature adopts a classification-based approach to defining virality (e.g., cascades smaller/larger than a given size; cascades doubling in a given time frame), these have limited use for applications that require comparison or prioritization among likely popular items. We focus here on approaches that provide \emph{real number predictions of popularity}.

Second, while low computational costs and memory constraints may not be prohibitive for offline or adhoc demonstration, scalability of this sort \feedback{can be especially relevant} when making predictions in real time. This is particularly true when evaluating large numbers of content items in parallel. To the best of our knowledge, only some previous work focused on the design of popularity prediction methods with the \emph{scalability as the main design goal} for applications in large-scale online platforms. Specifically, \cite{Tang17} proposed a method for video popularity prediction that uses a constant state per content item. Some prediction methods, e.g., Reinforced Poisson Process model \cite{Shen14} and SEISMIC~\cite{Zhao2015}, may be deemed to be computationally simple, but they still do not satisfy our target scalability constraints (see Sec.~\ref{sec:complexity} for details). Other methods, such as HIP~\cite{HIP17}, have scalability issues at prediction time and do not address the requirement of combining static and temporal features.

Third, flexibility in prediction time and the prediction time horizon are \feedback{desirable}. A simple approach to popularity prediction might involve a point-based prediction of growth up to a fixed content age based on features observable at a single point in time (e.g., content creation), but that model would not update in response to new information about the content (e.g., temporal features) and is limited in being able to predict for a fixed time horizon. \mvpolish{Supporting \emph{multiple time horizons} by using one model per horizon disallows predictions for previously unseen horizons.} Alternatively, many recent approaches to popularity prediction aim to provide estimates of total cascade size (at infinite time), limiting their utility for forecasting the ``urgency'' of cascade growth. Once again considering the content moderation application, queries about expected popularity may be made at multiple points in the content lifecycle from creation onward. In cases where content is removed---by platforms or by users themselves---cascades are truncated, making the evaluation of prediction accuracy for only a fixed prediction horizon difficult or impossible. Such truncated cascades are also unusable as training data in fixed or infinite horizon models.

Finally, the last \feedback{consideration} --- to \emph{leverage both static and temporal features} --- is important to ensure high prediction accuracy throughout the content lifecycle. Content views can result from complex interactions between content resharing and engagement, time zone use patterns, feed ranking algorithms, and organic features of the content and social network structure, so predictions benefit from insight into as much of this information as possible to the extent that the signals can be efficiently incorporated into the model. 
Clearly, any approach that relies only on event histories are likely to be inaccurate or unusable at early content ages. Similarly, approaches that rely only on static features will not adapt to new information provided by content engagement, resharing, and views.

In this paper, we propose a new popularity prediction model that (a) provides real number predictions, (b) has constant computation complexity and uses a \edit{small} space per content item, (c) can produce predictions for any given prediction time horizon specified at any given prediction time, and (d) leverages both static and temporal features. The model is based on a self-excited Hawkes point process model with exponentially decaying intensity, combined with prediction of model parameters by using both static and temporal features. This combination allows us to reduce the computation complexity of making predictions to constant time for any cascade size, but benefit from the analytically tractable estimators of the popularity over arbitrary future time horizons. 

Specifically, our basic prediction model uses only two point predictors, one for prediction of the number of points over a fixed reference time horizon (this is a hyper-parameter of the model) and one for the effective growth exponent which reflects the point process growth rate over time. This allows to use any point predictor developed and trained for making predictions for a specific time horizon, and then generalize this to support predictions for any given prediction time horizon by adding one extra point predictor. We also propose an extension that allows combining several point predictors of content view counts at different reference time horizons, increasing prediction accuracy while still using only a constant space per content item.

We demonstrate the accuracy and feasibility of our prediction method using a large-scale dataset of public Facebook posts over a two-month period. Our results demonstrate that high prediction accuracy can be achieved over different prediction time horizons, by using a few point predictors and that our models achieve performance that is comparable or better than a strong baseline that consists of using predictors designed and trained for specific prediction time horizons.

In Section~\ref{sec:related} we discuss related work. Section~\ref{sec:hawkes} lies down a framework for making predictions using self-excited point process models. Section~\ref{sec:predicting} defines our prediction models. \mvpolish{Section~\ref{sec:complexity} provides a discussion of our results}. Experimental results are presented in Section~\ref{sec:exp}. \mvpolish{In Section~\ref{sec:conc}, we provide concluding remarks. Appendix contains proofs and additional results.}

\section{Related work}
\label{sec:related}

Early work on predicting the popularity of online content considered various classification and regression models for fixed prediction time horizons using different \mvpolish{types of features} \cite{Szabo2010,CADKL14}. Much work has been devoted to understanding how information spreads in online social networks \cite{Gruhl2004,JL11,MSPLF12,ASHN13} and the role of social networks for information diffusion \cite{Bakshy2012,Cheng2018}. We refer the reader to surveys on web content popularity prediction \cite{MONIZ20191} and information cascade analysis \cite{Zhou21}. Models have been proposed for both popularity prediction (shares, views) and prediction of the number of users reached in an information cascade.  
We distinguish feature based models, generative models, and deep learning models, which we discuss in turn.

\paragraph{Feature based methods} Feature based prediction models use different types of features, including \emph{temporal features} (observation time, creation time, first view time), \emph{structural features} (cascade graph), \emph{user-item features}, and \emph{content features}. Several works considered prediction of an information cascade size by using information observed over an initial time period \cite{AH10,AH12,Kupavskii2012,CV12,Tsur2012,BKH13,Ma2013,BHK15}. Classification models \cite{Hong2011,Jenders2013,Kupavskii2012,Cui2013,CADKL14} and regression models \cite{Bakshy2009,Szabo2010,Kupavskii2012,Tsur2012} have been studied for prediction of information cascade sizes and prediction of occurrence of activity bursts \cite{Wang2015,CAKL16,SPL17}. Temporal features have been found to be important for prediction of content popularity and information sharing  \cite{Szabo2010,BKLD13,CADKL14}. Using network structural features is often not considered scalable \cite{SPL17}.

\paragraph{Generative models} Generative models assume events are generated according to a stochastic point process, which includes simple Poisson processes, survival analysis models, Hawkes point processes, and epidemic models. Different self-excited point process models have been used, including cascades of Poisson processes \cite{SJ10}, reinforced Poisson processes \cite{Shen14}, and Hawkes point processes and their variations \cite{Zhao2015, MRX16,KL16,HIP17}. Another class of models are multi-dimensional Hawkes processes, which allow to model different types of events and their mutual excitation \cite{Zhou13,Yang13,Zhou13b,Hui20}. Finally, epidemic models have also been used for modelling information diffusions \cite{RMKCX18,KRX20}. Most similarly, Hawkes point processes with exponentially decaying intensity were used for feature generation fed into a neural network predictor for predicting infinite-horizon watch time of Facebook videos \cite{Tang17}. Our work has similarities with these previous works in using a generative model and differs in emphasizing both scalability and making popularity predictions for arbitrary time horizons as the main design goals.

\paragraph{Deep learning models} Deep learning models use neural networks as prediction models or for learning numerical vector representation (embeddings) of temporal or structural features for popularity prediction. Several works extended self-excited point process models with neural networks, including DeepHawkes~\cite{DeepHawkes17}, NeuralHawkes~\cite{Mei2017}, and SIR-Hawkes~\cite{RMKCX18}. Neural networks have been used for representations of event histories \cite{Du2016}, incidence curves \cite{ZL18}, information diffusion networks~\cite{LMGM17,Lamprier2019}, fusion of content and temporal features~\cite{Liao19}, representations of structural and temporal information~\cite{Chen19}, and social network interactions~\cite{Cao20}. Deep learning based models for popularity prediction are not scalable for our intended scenarios, as they typically require inputs that grow linearly in the number of past events and are complex or expensive to use for making predictions over arbitrary time horizons. 

\section{Methodology}
\label{sec:hawkes}

\mvrev{In this section, we first present some results on self-excited point processes in Section~\ref{sec:formulation}, which are used to define our prediction method in Section~\ref{sec:predicting}.}
\subsection{Self-excited point processes}
\label{sec:formulation}

\subsubsection{Background}

\mvrev{We consider generative models defined as point processes, with points representing occurrance times of view events of a content item.} A realization of a \emph{point process} on $\reals_+$ is a sequence of points $0\leq T_1\leq T_2 \leq \cdots$ that can be equivalently represented by a counting variable $N(t)$ defined as the number of points in $[0,t)$, i.e. $N(t) = \sum_{i\geq 1} \mathbf{1}_{\{0\leq T_i < t\}}$, for any $t\in \reals_+$. A \mvrev{\emph{stochastic point process}} has the \emph{stochastic intensity function} defined by
$$
\lambda(t) = \lim_{\epsilon \downarrow 0} \frac{\EC{N(t+\epsilon) - N(t)}{\mathcal{F}_t}}{\epsilon},
$$

\noindent where $\mathcal{F}_t$ is the history of the point process up to time $t$. \mvrev{Intuitively, we can think of $\lambda(t)$ as 
the conditional probability that there is a point in $[t,t+\epsilon)$, conditional on the history ${\mathcal F}_t$, for small $\epsilon$.}

A \emph{Hawkes point process} is defined by the stochastic intensity function 
$$
\lambda(t) = \lambda_0(t) + \sum_{i=1}^\infty \phi_{Y_i}(t-T_i) \mathbf{1}_{\{0\leq T_i < t\}},
$$ 
where $\lambda_0$ and $\phi_y$ are given functions and $y \in \reals_+$.  Here $Y_0, Y_1, \ldots$ are assumed to be independent and identically distributed random variables (referred to as \emph{marks}) according to distribution $F_Y$, which are independent of the points $T_1, T_2,\ldots$. Following standard definition, we assume that $\phi_y(x)$ is of the form \mvrev{$\phi_y(x) = y \phi(x)$,}
where $\phi(x)$ is a \emph{kernel function}. Under this assumption, $Y_i$ is the size of a jump in the stochastic intensity function.

Let $\mu$ be \mvrev{the expected contribution of a point to the value of the stochastic intensity function} defined by
\mvrev{$\mu = \E_{Y\sim F_Y}\left[\int_0^\infty \phi_Y(t)dt\right]$.}
We assume that $\mu < 1$, which ensures stability of the point process. 

The framework of self-excited point processes accommodates different instances of \mvrev{stochastic point} processes. Here we consider two notable examples.

\paragraph{Exponentially decaying kernel} The Hawkes point process \emph{with exponentially decaying intensity} is defined by the kernel function
\begin{equation}
\phi(x) = e^{-\beta x},
\label{equ:phiyexp}
\end{equation}
where $\beta > 0$ is a parameter and assuming that $\lambda_0(t) = \lambda(0)\phi(t)$\mvrev{, for some initial value $\lambda(0)>0$.}  In this case, we have
$$
\lambda(t) = \lambda(0)e^{-\beta t} + \sum_{i=1}^\infty Y_i e^{-\beta(t-T_i)} \mathbf{1}_{\{0\leq T_i < t\}}.
$$

We will use the change of variable such that $Y_i = \beta Z_i$ for a random variable $Z_i$ with distribution $G$. We may interpret $Z_i$ as a population size (neighbors of a node in a social network) and $\beta$ as a rate parameter (rate of interactions between nodes in a social network). Let $\rho_r$ denote the $r$-th moment of $Z_i$, i.e.
\mvrev{
$\rho_r = \int_0^\infty z^r dG(z)$.
Note that $\E[Y_1] = \beta\rho_1$ and $\mu = \rho_1$.} 

We will later discuss that Hawkes point processes with exponentially decaying intensity have certain desirable properties for scalable popularity prediction over arbitrary time horizons. 

\paragraph{Power-law decaying kernel} Another commonly used kernel function is the \emph{power-law kernel} defined as
\begin{equation}
\phi(x) = \left\{
\begin{array}{ll}
\phi(0) & \hbox{ if } 0 \leq x \leq \tau,\\
\phi(0)\left(\frac{\tau}{x}\right)^{1+\theta} & \hbox{ if } x > \tau,
\end{array}
\right . 
\label{equ:power}
\end{equation}
where $\phi(0) > 0$, $\tau > 0$ and $\theta > 0$ are parameters. For instance, this kernel was used in \cite{Zhao2015} \mvrev{and \cite{HIP17}} to model information cascades. 

The framework presented in this section has the following interpretation in the context of popularity prediction of content items. We may interpret each point as a content view event that excites subsequent content view events. \mvrev{For the Hawkes point process with exponentially decaying kernel}, the random variable $Z_i$ can be interpreted as the number of potential users that can be reached resulting from \mvrev{the content view event at time $T_i$}. The parameter $\beta$ is the rate at which users consume content. The parameter $\mu$ is the expected number of subsequent content view events triggered by a content view event. The kernel function models the time-decay of the stochastic intensity function \edit{components triggered} by content view events, \mvrev{capturing their diminishing influence over time}.

\subsubsection{Counts over future time horizons}
\label{sec:moments}

For popularity prediction for a content item, we are interested in predicting the number of content view events over a given time horizon at a given prediction time, having observed the history of the content views up to the prediction time. Using the framework introduced in previous section, given a prediction time $s$ and a time horizon up to time instance $t > s$, we are interested in predicting the value of $N(t)-N(s)$, having observed the history $\mathcal{F}_s$. 

\paragraph{Infinite time horizon} For any stable Hawkes point process, the conditional expected number of points over an \emph{infinite} time horizon originating at a time instance $s\geq 0$, conditional on the history ${\mathcal F}_s$, is given as
\begin{equation}
\lim_{t\rightarrow\infty} \EC{N(t)-N(s) }{\mathcal{F}_s} = \frac{1}{1-\mv{\mu}} \lim_{t\rightarrow \infty}\Lambda(s,t)
\label{equ:genlimexp}
\end{equation}
where 
$$
\Lambda(s,t) = \Lambda_0(t)-\Lambda_0(s) +  \sum_{i\geq 1} y_i\left(\Phi(t-T_i)-\Phi(s-T_i)\right){\mathbf 1}_{\{0\leq T_i < s\}},
$$
and $\Lambda_0$ and $\Phi$ are the primitive functions of $\lambda_0$ and $\phi$, respectively, \mvrev{i.e. $\Lambda_0(x) := \int_0^x \lambda_0(u)du$ and $\Phi(x):=\int_0^x \phi(u)du$}. Here $\Lambda(s,t)$ 
is the conditional expected number of points in $[s,t]$, induced by the intensity function $\lambda_0$ and the intensity function \mvrev{components} excited by points in $[0,s]$, conditional on the history $\mathcal{F}_s$.

For the Hawkes point process with exponentially decaying intensity, the expression in (\ref{equ:genlimexp}) boils down to
\begin{equation}
\lim_{t\rightarrow\infty} \EC{N(t)-N(s)}{\mathcal{F}_s} = \frac{1}{\alpha}\lambda(s)
\label{equ:limexp}
\end{equation}
where $\alpha = \beta(1-\rho_1)$. For the reasons explained shortly, we refer to $\alpha$ as \emph{the effective growth exponent}. Note that (\ref{equ:limexp}) is a function only of the intensity $\lambda(s)$ and the effective growth exponent $\alpha$. 

\paragraph{Arbitrary time horizons} It is not tractable to have an explicit formula for the conditional expected count over an \emph{arbitrary} time horizon\mvrev{---which is our objective---}for all Hawkes point processes. However, we offer the following bounds. 

\begin{proposition} For any stable Hawkes point process, for every $0\leq s\leq t$, we have
$$
\Lambda(s,t) \leq \EC{N(t)-N(s)}{{\mathcal F}_s} \leq \frac{1}{1-\mv{\mu}}\Lambda(s,t).
$$
\label{pro:bounds}
\end{proposition}

Proof of this proposition is given in Appendix~\ref{sec:bounds}. Note that for any fixed value of \mv{$\mu < 1$}, 
$\EC{N(t)-N(s)}{\mathcal{F}_s}$ is within a constant factor of $\Lambda(s,t)$. Intuitively, the bounds in Proposition~\ref{pro:bounds} are tighter the nearer the value of 
\mv{$\mu$} is to zero (small expected number of points excited by a point).

\paragraph{Arbitrary time horizons for exponential kernel}
For the Hawkes point process with exponentially decaying intensity, we can characterize the conditional expected number of points over an \emph{arbitrary} time horizon, conditional on the observed history up to a time instance, as stated in the following proposition. \mvrev{This is a key proposition for defining our prediction model in Section~\ref{sec:predicting}.}

\begin{proposition} For the Hawkes point process with exponentially decaying intensity, for every $0 \leq s \leq t$, we have
\begin{equation}
\EC{N(t) - N(s)}{ \mathcal{F}_s} = \frac{1}{\alpha}\left(1-e^{-\alpha(t-s)}\right)\lambda(s).
\label{equ:expn}
\end{equation}
\label{prop:exp}
\end{proposition}

Proof is given in Appendix~\ref{app:exp}. From (\ref{equ:expn}), observe that the conditional expected count of points converges exponentially 
to its limit value with rate $\alpha$, which provides a justification for referring to $\alpha$ as the effective growth exponent.  

The effective growth exponent $\alpha$ admits the following intuitive interpretation. Note that we can write (\ref{equ:expn}) as
$$
\EC{N(t)-N(s)}{\mathcal{F}_s} = \EC{N(+\infty)-N(s)}{\mathcal{F}_s}(1-e^{-\alpha(t-s)}).
$$
For any given $\gamma \in (0,1)$, let $\tau_\gamma$ be the length of the time horizon at which the conditional expected count is equal to factor $\gamma$ of its limit value. It is easy to derive that
\begin{equation}
\tau_\gamma = c_\gamma\frac{1}{\alpha},
\label{equ:taugamma}
\end{equation}
with constant $c_\gamma = \log(1/(1-\gamma))$. Hence, we can interpret the reciprocal value of $\alpha$ as a \emph{characteristic time}. 

A notable property of Hawkes point processes with exponentially decaying intensity is that $\Lambda(s,t)$ and $\E[N(t)-N(s)\mid \mathcal{F}_s]$ depend on the history $\mathcal{F}_s$ only through the value of the stochastic intensity $\lambda(s)$ at time instance $s$. This can be leveraged for making scalable predictions by using low-complexity estimators of $\lambda(s)$. This stands in contrast to other Hawkes point processes, which require using more expensive computations.

\subsection{Prediction method}
\label{sec:predicting}
In this section we present our model for predicting popularity of social media items over arbitrary time horizons. The model is designed with \emph{scalability} as the main design requirement. The  idea behind our approach is to use a Hawkes model with parameters determined by a learned mapping between a vector representation of the content features and point process parameters. This approach allows us to reduce the computation complexity of making predictions to constant time with respect to the observed events in the cascade $N(s)$, \mvpolish{and} benefit from the analytically tractable estimators of the popularity over arbitrary future time horizons. 

\subsubsection{Prediction model}

The model is based on \mvrev{the following expression for the conditional expected number of points up to future time $s+\delta$, for given prediction time $s$ and prediction time horizon $\delta \geq 0$, and an arbitrarily fixed \emph{reference horizon} $\delta^* > 0$,} 
\begin{equation*}
	\EC{N(s+\delta)}{\mathcal{F}_s} = N(s) + \frac{1-e^{-\alpha \delta}}{1-e^{-\alpha \delta^*}}\left(\EC{N(s+\delta^*) }{ \mathcal{F}_s} - N(s)\right)
	\label{equ:expn2}
\end{equation*}
\mvrev{which follows from Proposition~\ref{prop:exp}.} 

\mvrev{The expression above has two unknown parameters: (a) the conditional expected number of points at the reference time horizon, $\EC{N(s+\delta^*)}{\mathcal{F}_s}$, and (b) the effective growth exponent $\alpha$. These unknown parameters need to be inferred for any given features of a content item by using training data}. 

\mvrev{
Let $\hat{N}(\delta; s)$ denote the predictor of $N(s+\delta)$ given history $\mathcal{F}_s$ and $\hat{\alpha}$ denote the predictor of $\alpha$. Let us also use a logarithmic transformation of the prediction variable by defining $Y(\delta;s) = \log(\hat{N}(\delta; s) - N(s))$.} Then, we can write
\begin{equation}
Y(\delta;s) = Y(\delta^*;s) + \log\left(\frac{1-e^{-\hat{\alpha}\delta}}{1-e^{-\hat{\alpha} \delta^*}}\right)
\label{equ:y}
\end{equation}

with $Y(\delta^*;s)$ and $\hat{\alpha}$ being values of two predictors defined as follows. The first predictor is for the log-transformed number of points over the reference time horizon,
\begin{equation}
Y(\delta^*;s) = f(x, \tau(\mathcal{F}_s); \theta),
\label{equ:ystar}
\end{equation}

\noindent where $x$ is \mvrev{the vector of static features and $\tau(\mathcal{F}_s)$ is the vector of temporal features derived from $\mathcal{F}_s$ of the content item, and $\theta$ is the regression model parameter.} The second predictor is for the effective growth exponent:
\begin{equation}
\hat{\alpha} = g(x, \tau(\mathcal{F}_s); \theta'),
\label{equ:hatalpha}
\end{equation}

\noindent where $\theta'$ \mvrev{is the regression model parameter}. We use temporal features $\tau(\mathcal{F}_s)$ that can be computed in constant time, \mvrev{which is required by our scalability requirement}. 

\mvrev{In summary, our prediction method amounts to predicting popularity of a content item at time $s+\delta$, at prediction time $s$, and any given prediction time horizon $\delta$, by using equation (\ref{equ:y}) with $Y(\delta^*;s)$ and $\hat{\alpha}$ defined by functions of the static feature vector $x$ and the temporal feature vector $\tau(\mathcal{F}_s)$ as given in (\ref{equ:ystar}) and (\ref{equ:hatalpha}), respectively.}

\subsubsection{Training details}
\mvrev{Functions $f$ and $g$, in (\ref{equ:ystar}) and (\ref{equ:hatalpha}) respectively, can be implemented by using \mvpolish{standard machine learning algorithms}. In our evaluations in Section~\ref{sec:exp} we used gradient boosted decision trees, trained independently for $f$ and $g$. For training parameters of $f$, we use $(x_i,y_i)$ as training examples where $x_i$ is the vector of static and temporal features and $y_i$ is the number of points observed over the reference time horizon for training example $i$. Similarly, for training parameters of $g$, we use $(x_i,y_i)$ as training examples with $x_i$ defined as before and $y_i$ defined to be an estimate of the effective growth exponent for content item $i$. We discuss estimators of the effective growth exponent in Section~\ref{sec:estexp}.}

\mvrev{A notable property of our prediction model is that it requires using only two point predictors, while allowing for making predictions for any given prediction time horizon. With scalability in mind, we consider point predictors which can be computed in constant time with respect to the observed history of cascade. Notice that the predicted value for the length of prediction horizon $\delta = \delta^*$ is equal to $Y(\delta^*; s)$. In this case, our predictor is guaranteed to be as accurate as the predictor optimized for the reference time horizon $\delta^*$. For $\delta \neq \delta^*$, the predictor may have a worse accuracy than a predictor optimized for the time horizon $\delta$. We will evaluate this empirically in Section~\ref{sec:exp}, where we will see that the proposed method can achieve competitive performance to predictors optimized for specific prediction time horizons.}

\subsubsection{Combining multiple point predictors}
\label{sec:multiple}

\mvrev{We can extend our prediction method to using one or more point predictors, which may increase prediction accuracy.} Let $\hat{N}(\delta^*_1; s)$, $\ldots$, $\hat{N}(\delta^*_m; s)$ be point predictors for given values of reference horizons $\delta_1^* < \delta_2^* < \cdots < \delta_m^*$\mvrev{, for some given $m\geq 1$}. The prediction method is defined by combining outputs of these point predictors.

We consider two different predictors that combine outputs of point predictors by using different combining functions. 

\paragraph{Arithmetic mean aggregation} The first predictor combines outputs of different point predictors ($\hat{N}(\delta_1^*;s), \ldots, \hat{N}(\delta_m^*;s)$) by using the arithmetic mean aggregation,
which amounts to the following predictor for the log-transformed prediction variable:
\begin{equation*}
Y(\delta;s) = \log\left(\frac{1}{m}\sum_{i=1}^m \frac{1}{1-e^{-\hat{\alpha}\delta^*_i}}e^{Y(\delta_i^*; s)}\right) + \log\left(1-e^{-\hat{\alpha}\delta}\right).
\end{equation*}

\paragraph{Geometric mean aggregation} The second predictor combines outputs of point predictors ($\hat{N}(\delta_1^*;s), \ldots, \hat{N}(\delta_m^*;s)$) by using the geometric mean aggregation,
which amounts to the following predictor for the log-transformed prediction variable:
\begin{equation}
Y(\delta;s)  = \frac{1}{m} \sum_{i=1}^m Y(\delta_i^*;s) + \log\left(\frac{1-e^{-\hat{\alpha}\delta}}{\left(\prod_{i=1}^m \left(1-e^{-\hat{\alpha}\delta_i^*}\right)\right)^{1/m}}\right).
\label{equ:y_m}
\end{equation}

\noindent We will evaluate the accuracy of prediction models with one or more point predictors in Section~\ref{sec:exp}.

\subsubsection{Estimating the effective growth exponent} 
\label{sec:estexp}

\mvrev{To train the predictor of the effective growth exponent in equation (\ref{equ:hatalpha}), we need training examples with the response variable corresponding to the effective growth exponent. One way to compute the effective growth exponent is to use MLE for given observed data. This is computationally expensive so we discuss two simpler estimators.}

\paragraph{Mean value based estimator} By Proposition~\ref{prop:exp}, for every $t\geq 0$,
$$
\EC{N(+\infty)-N(t) }{ \mathcal{F}_t} = \frac{\lambda(t)}{\alpha}.
$$

We can show that 
$$
\EC{\int_s^\infty (N(+\infty)-N(t))dt }{\mathcal{F}_s} = \frac{1}{\alpha} \EC{N(+\infty)-N(s)}{\mathcal{F}_s}
$$ which \mvrev{follows from derivations in Appendix~\ref{sec:meanalpha}.} This leads us to define the following estimator
$$
\hat{\alpha} = \frac{N(+\infty)-N(s)}{\int_s^\infty (N(+\infty)-N(u))du}.
$$

Suppose $s = 0$ and $N(s) = 0$ and let $T_1, T_2, \ldots, T_n$ denote the observed points. It can be shown that 
$$
\int_0^\infty (n-N(t))dt =  \sum_{i=1}^n T_i
$$

\noindent which follows by some simple calculations provided in Appendix~\ref{sec:meanalpha} \mvpolish{\cite{haimovich2020scalable}}. Hence, we have
$$
\hat{\alpha} = \frac{1}{\frac{1}{n}\sum_{i=1}^n T_i}.
$$
\noindent \mvrev{This shows that $\hat{\alpha}$ is the reciprocal of the mean point time.} 

\paragraph{Quantile value based estimator} An alternative estimator can be defined based on computing a quantile value as described next. For any fixed value $\gamma \in (0,1)$, let
$$
T_\gamma = \inf\{t > 0 : N(t) \geq \gamma N(+\infty)\}.
$$

\noindent Notice that if $\gamma = 1/2$, then we can interpret $T_{1/2}$ as the median value of the observed point times. Intuitively, we may think of $T_\gamma$ as of an estimator of $\tau_\gamma$, defined by $\EC{N(\tau_\gamma) }{ \mathcal{F}_0} = \gamma\EC{  N(+\infty)}{ \mathcal{F}_0}$. We already noted in (\ref{equ:taugamma}) that $\tau_\gamma = \log(1/(1-\gamma))/\alpha$. Hence, this leads us to define $\hat{\alpha} = 1/T_\gamma$, provided that $T_\gamma > 0$.

In Appendix~\ref{sec:quantileerror}, we provide a theoretical bound on the bias of the quantile value based estimator. In Section \ref{sec:exp}, we empirically compare the two estimators on real-world data.

\section{Discussion}
\label{sec:complexity}

\mvrev{In this section we discuss \mvpolish{the} computation complexity of some previously proposed methods based on point process models as well as of our prediction method presented in Section~\ref{sec:predicting}.} 

In order to make predictions by using expressions for $\E[N(t)-N(s)\mid \mathcal{F}_s]$ or $\Lambda(s,t)$ \mvrev{discussed in Section~\ref{sec:formulation}} for different point process models, we need to compute these values which has certain computation cost. This computation cost is incurred both at \emph{training time} (for computing values of prediction variables used for supervised learning) and at \emph{prediction time}. Moreover, additional computation cost is incurred for estimating unknown model parameters \mvrev{at training time}. 

For general Hawkes point processes, the computation of $\E[N(t)-N(s)\mid \mathcal{F}_s]$ or $\Lambda(s,t)$ can be prohibitively expensive for implementation in large-scale online platforms. Evaluation of these quantities have $\Omega(N(s))$ computation complexity, i.e. it is at least linear in the number of points in the observed history. For popularity prediction in social media platforms, this number can be large, in the order of millions and possibly even larger. 

We next discuss computation complexity of these evaluations for several well-known methods (namely, \mvrev{Reinforced Poisson Process}, SEISMIC, Hawkes Intensity Process, and Hawkes with exponential kernel). We do not \mvrev{discuss computation complexity of}  deep learning extensions of these models \mvrev{as they have same or higher} complexity. 

\paragraph{Reinforced Poisson Processes} RPP model \cite{Shen14} has the stochastic intensity function $\lambda(t) = p f(t) N(t)$ where $p$ is a positive-valued infection-rate parameter and $f(t)$ is a probability density function. The model assumes $f$ to be a log-normal density function, which has two parameters. This model does not exactly fall in the framework of Hawkes point processes, but it is a self-excited point process model. The conditional expected number of points over an arbitrary time horizon is given by
$$
\E[N(t)-N(s)\mid \mathcal{F}_s] = N(s) \left(e^{p(F(t)-F(s))}-1\right).
$$

\noindent The model requires to track the total number of points observed by any given time, which can be efficiently tracked in a streaming computation setting. However, the model is computationally expensive as it requires to fit model parameters for each content item using a Maximum Likelihood Estimator (MLE), which requires using an iterative optimization method. Specifically, the time complexity of this approach $\Omega(M\times N(s))$ is proportional to the number of iterations $M$ of the optimization method (which can be considerably large in practice) times the number of points in the history $N(s)$.  

\paragraph{SEISMIC} This model \cite{Zhao2015} is a Hawkes point process model with a power-law kernel $p\phi(x)$ where $\phi(x)$ is given in (\ref{equ:power}). The model is defined by letting marks $Y_i$ be the degrees $d_i$ of nodes re-sharing information in an online social network. The two parameters of the function $\phi(x)$ are assumed to be hyper-parameters, and an MLE is used to estimate parameter $p$ by using the observed part of a cascade. This estimator can be expressed in a closed form as 
$$
\hat{p} = \frac{N(s)}{\sum_{i=1}^{N(s)} d_i \Phi(s-T_i)}.
$$

The paper \cite{Zhao2015} uses a variant of this estimator that involves some smoothing. \mvrev{Clearly,} the computation complexity for evaluating the value of estimator $\hat{p}$ is $\Omega(N(s))$.

\mvrev{\paragraph{Hawkes Intensity Process} The HIP method \cite{HIP17} assumes a Hawkes point process with a power-law kernel function and is based on estimating the model parameters by fitting the expected value of the stochastic intensity function to observed data at fixed time instances. For general Hawkes point processes, the expected value of the stochastic intensity function obeys a convolutional equation, which is leveraged by the proposed approach. This approach still requires using an iterative optimization method and has the time complexity comparable to RPP.}

\paragraph{Hawkes with exponential kernel} 
\mvrev{For the Hawkes point process model with exponentially decaying intensity, by Proposition~\ref{prop:exp}, we need to evaluate the value of the stochastic intensity $\lambda(s)$ in order to compute the value of $\EC{N(t) - N(s)}{ \mathcal{F}_s}$.} \mvrev{The stochastic intensity $\lambda(s)$} can be approximated by a \emph{velocity statistic} which measures the local rate of points at time $s$. For instance, we may define the velocity as the rate of points observed over $[s-d,s]$ for some fixed value $d>0$. Velocity can be efficiently tracked and queried in constant time by using a sliding-window algorithm over the stream of observed points~\cite{datar2002maintaining}. For estimating the other two parameters of the model, namely $\rho_1$ and $\beta$, one may use an MLE optimization method. This approach, as in the methods mentioned above, may induce significant computation costs. An alternative approach is to use an estimator for the effective growth exponent $\alpha$. This parameter is both sufficient for prediction purposes (see Proposition~(\ref{prop:exp})) and \mvrev{an estimator of this parameter} be efficiently computed (see Section~\ref{sec:estexp}). 

\section{Experimental results}
\label{sec:exp}

In this section we present our numerical results. We first provide basic information about datasets that we used for training and evaluation, the models we chose for comparison and our evaluation metrics. We then provide results on the accuracy of predictions over infinite and then varied time horizons. Our choice of baseline models includes previously proposed popularity prediction models based on self-excited point processes, and separately trained machine learning models for specific prediction time horizons. Overall our results show that our proposed method can provide more accurate predictions than other self-excited point process models, and that our method achieves competitive performance to models trained for specific prediction horizons.  

\subsection{Datasets, models and evaluation metrics}

\paragraph{Datasets.} \edit{For our experiments, we used datasets containing de-identified public Facebook posts created by pages (Facebook accounts of companies, brands, celebrities, and other public entities) and collected over different time periods. These datasets cover a large number of view and reshare events  -- in the order of billions -- and hundreds of thousands of posts. Specifically, we used a dataset containing 100 thousand public page posts which were reviewed by moderators but deemed to not violate Facebook Community Standards. These posts were created within 2 weeks in October 2020; we tracked their reshares and views for up to 2 months after creation. The number of views recorded on these posts is in the order of hundreds of billions. We also used a second dataset containing 200 thousand randomly sampled public page posts created within 1 week in November 2019, and also tracked their reshares and views for up to 2 months after creation, collecting timestamps of several billions of such events. We used the first dataset to evaluate prediction accuracy of different models for infinite horizons. For validating performance on the varied prediction horizons we used both datasets and obtained similar results. Hence, for varied prediction horizons we only present results on the second dataset. \mvrev{We believe that datasets we use are typical and hence the claims made in this section would generalize to other datasets.}}  

\paragraph{Our prediction model}
\label{sec:prediction_model_details} The Hawkes model we propose is defined in Section~\ref{sec:predicting}. We use gradient boosted decision trees from the scikit-learn library~\cite{friedman2002stochastic} for point predictors of the view counts for given reference horizons and the effective growth parameters. We use a set of 1889 features, which could be categorized into \emph{content features} (properties of the post), \emph{page features} (properties of the account that created the post), and \emph{engagement features} (patterns of users' interactions with the post and the page). Appendix~\ref{app:features} provides details on these groups of features and their cumulative importance for both regressors. As expected, engagement features have the highest importance scores for both regressors. However, the long term patterns of a cascade's growth -- as indicated in the case of predicting effective growth exponent $\alpha$ -- are better explained by the characteristics of the page and the \emph{page-level engagement} features. In contrast, the \emph{content engagement} features are by far the most important for popularity prediction over shorter horizons. 

\paragraph{Baselines} We compare prediction accuracy of our model against several carefully chosen baselines drawn from \edit{relevant} literature. Our first set of baselines are taken from the class of generative models based on self-excited point processes. In particular, we compare against a variant of SEISMIC \cite{Zhao2015} adapted for predicting popularity of Facebook posts following \cite{Tang17} and the RPP model \cite{Shen14}. \dk{We used the source code of SEISMIC model from the original paper~\footnote{http://snap.stanford.edu/seismic/}. For RPP, we have not found the original source code of the model and opted for reproducing it in Python.} These baseline models are representative of the family of generative models based on self-excited point processes, and their computation complexity is not so high as to make them unusable on our data (in contrast with other more complex models like those that combine deep learning with self-excited processes). Our second set of baselines consists of prediction models separately trained for specific reference prediction horizons (hereafter ``PB''), and a prediction model that uses the the horizon as the feature (hereafter ``HF''). We will provide some more discussion about the baseline models in the following sections.

\paragraph{Evaluation metrics} Following~\cite{Zhao2015}, we evaluated prediction accuracy using Median Absolute Percentage Error (MAPE) and $\tau$ Rank Correlation; we also added some evaluation results using Root Mean Squared Error (RMSE). 

\begin{center}
\begin{table}[t]
		\caption{Prediction performance for the proposed Hawkes model vs. SEISMIC-CF, overall, and conditional on content popularity (Low, High) or prediction time (Early, Late).}
		\label{exp:tab:infinite}
\begin{tabular}{p{0.5in} c c c c c c }
\hline
\multirow{2}{*}{Dataset} & \multicolumn{3}{c }{Hawkes}       & \multicolumn{3}{c }{SEISMIC-CF}      \\ \cline{2-7} 
                         & MAPE & $\tau$ & RMSE             & MAPE & $\tau$ & RMSE             \\ \hline
Overall                    & 0.565 & 0.821  & 2.0e6 & 0.698 & 0.769  & 6.5e6 \\ \hline
Low           & 0.651 & 0.713  & 9.8e4 & 0.802 & 0.633  & 1.6e7 \\ 
High           & 0.552 & 0.796  & 2.2e6& 0.685 & 0.744  & 2.4e7 \\ \hline
Early       & 0.451 & 0.824  & 1.4e6 & 0.667 & 0.752  & 9.9e7 \\ 
Late        & 0.573 & 0.821  & 2.3e6 & 0.737 & 0.762  & 2.8e7 \\ \hline
\end{tabular}
\end{table} 
\end{center}

\subsection{Predictions for infinite horizons}
\label{sec:infinite}

In this section we present our numerical results on the prediction accuracy of our model and compare with two baselines, namely, a variant of SEISMIC and RPP models, which we introduced in Section~\ref{sec:complexity}. The presented numerical results demonstrate that our model can achieve superior prediction accuracy than these baseline models, by leveraging static features, and that this can be achieved at a much smaller computation time cost.   

We compare our approach against a SEISMIC-CF variant of the model proposed for Facebook cascades in \cite{Tang17}. \edit{We used default values for the constant node degree parameter proposed for SEISMIC-CF and for the kernel function parameters. We have explored various other settings of parameter values and obtained similar results.} As it can be seen in Table~\ref{exp:tab:infinite}, our model outperforms SEISMIC-CF on both Median APE and Rank Correlation by a margin of $13\%$ and $5\%$, respectively. This also holds true across different splits we tested on -- namely, low vs. high popularity items (less or more than 1000 views) and early vs. late predictions with respect to content age at prediction time (less or more than 24 hours since content creation). The performance gap is especially striking when comparing predictions by using the RMSE metric, where for low popularity items and early predictions, our model is orders of magnitudes more accurate than SEISMIC-CF. 

We have also conducted experiments to compare against RPP, which we introduced in Section \ref{sec:complexity}. As discussed in Section~\ref{sec:complexity}, the computation complexity of fitting RPP model for each content item is proportional to the product of the number of steps of the MLE optimization algorithm and the number points in the observed history. In out settings, this was in the order of minutes for high popularity content items in comparison to less than a second for our proposed model. We managed to evaluate RPP on a small subset of content items in our dataset and achieved a MAPE of \edit{$4.1$}, which is significantly worse than for our model. 
\begin{figure}
	\begin{center}
		\includegraphics[width=1.0\linewidth]{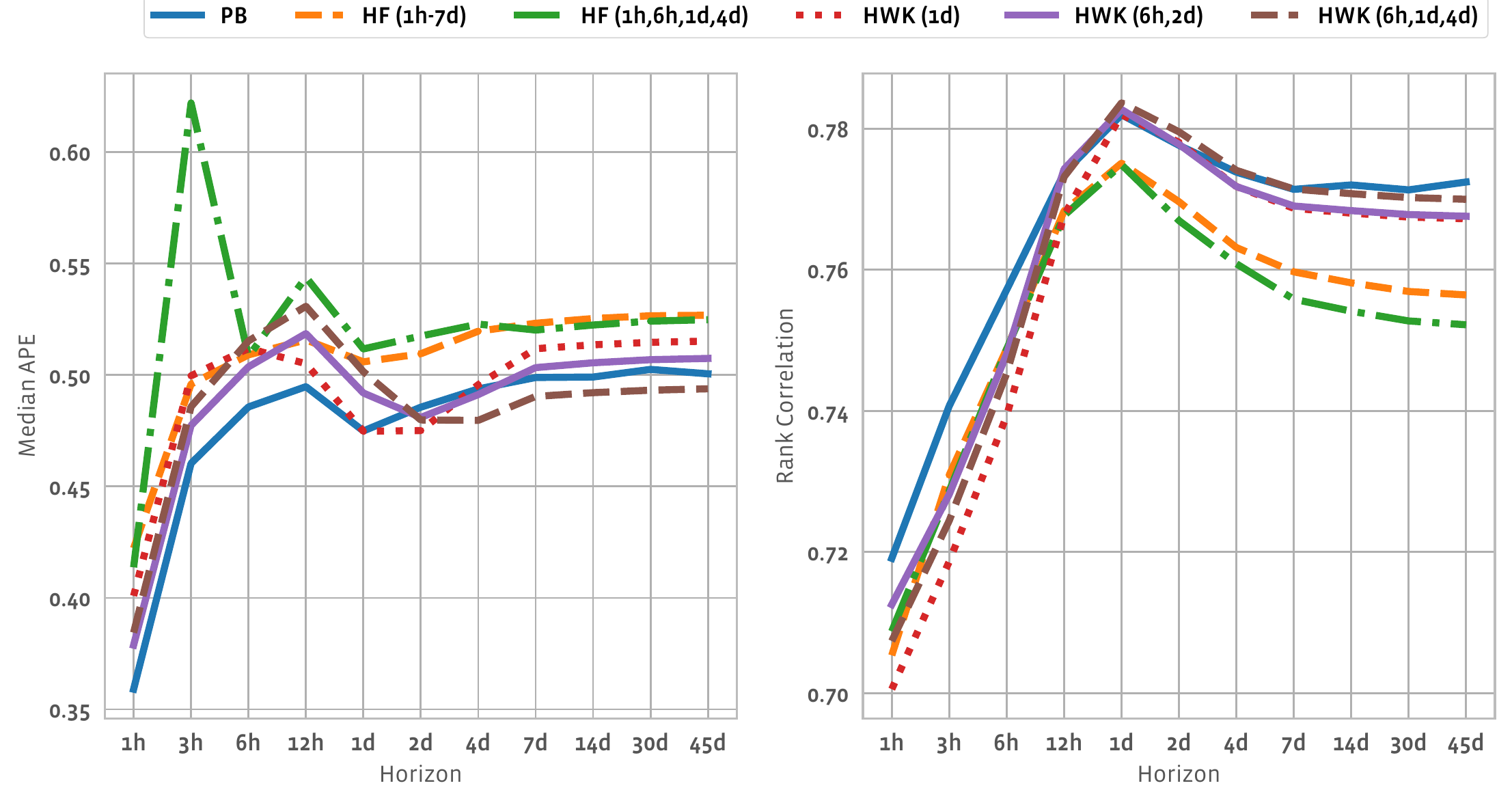}
		\caption{\mvpolish{Prediction performance for different horizons: (left) median absolute prediction error and (right) rank correlation. The results are for the proposed Hawkes models with one reference time horizon (HWK (1d)), two reference time horizons (HWK (6h,4d)) and three reference time horizons (HWK (6h,1d,4d)), point-based models (PB), and horizon-as-feature models, one trained on all considered horizons between 1 hour and 7 days (HF (1h-7d)) and another one trained only on a subset of them (HF (1h,6h,1d,4d)).}}
		\label{fig:prediction_overall}
	\end{center}
\end{figure}

\subsection{Predictions over arbitrary horizons}

In this section we compare prediction performance of our model against two different baseline models, including models that are separately trained for specific prediction time horizons and models that use the prediction time horizon as the input feature. More specifically, we consider: \mvrev{(a) \emph{Point-based (PB)} models that are trained separately for every given prediction time horizon. Although in practice it might not be feasible to maintain a family of models for potentially infinite horizons of interest, this approach provides a good estimate for upper bound performance when a dedicated model is trained for each horizon. (b) \emph{Horizon-as-feature (HF)} models for popularity prediction that use the prediction time horizon as the input feature. This requires training examples sampled at a multitude of horizons $\delta$, i.e. 
\mvrev{$
Y(\delta;s) = h(\delta, x, \tau(\mathcal{F}_s); \theta)
$, 
which has an additional independent variable $\delta$.} }

The PB models may be regarded as a strong baseline for comparison of prediction performance for specific prediction time horizons, as they are trained for these specific prediction time horizons. The HF models may be regarded as a natural class of prediction models.

For training HF models, we sample prediction time horizons in the range between $1$h and $7$d for each content item, hence synthetically increasing the size of the training set by the number of considered horizons, i.e., eight-fold for a model variant trained on all considered horizons in the range (HF (1h-7d)) and four-fold for a model variant trained only on a subset of them (HF (1h,6h,1d,4d)).

We compare the performance of our model against the aforementioned baselines for different reference time horizons of our model. We denote our model as HWK($\delta_1^*,\ldots, \delta_m^*$) for given reference prediction time horizons $\delta_1^*, \ldots, \delta_m^*$.


\mvrev{As seen in Figure~\ref{fig:prediction_overall}, all considered Hawkes models outperform the HF baselines on longer horizons ($\delta > 24$h) with the best one (HWK (6h,1d,4d)) having an average decrease of $7 \%$ in Median APE and an average increase of $2\%$  in Rank Correlation. Evidently, the HF model struggles to generalize beyond the horizons it has been trained on, as seen from the sharp drops of the HF (1h,6h,1d,4d)'s performance for $\delta = 3$h, $12$h, $2$d in comparison to the HF (1h-7d) variant trained on all horizons in the range. Last but not least, our model also reaches a parity in performance with PB models for $\delta > 24$h, suggesting its good generalization capability for long prediction horizons.}

\dkpolish{We further discuss tuning of the reference horizon parameters $\delta$ and performance of the models on cascades of different sizes in Appendix~\ref{sec:tuning} and Appendix~\ref{sec:performance_cascade_size}, \mvpolish{respectively}.} 

\begin{figure}
	\begin{center}
		\includegraphics[width=0.47\linewidth]{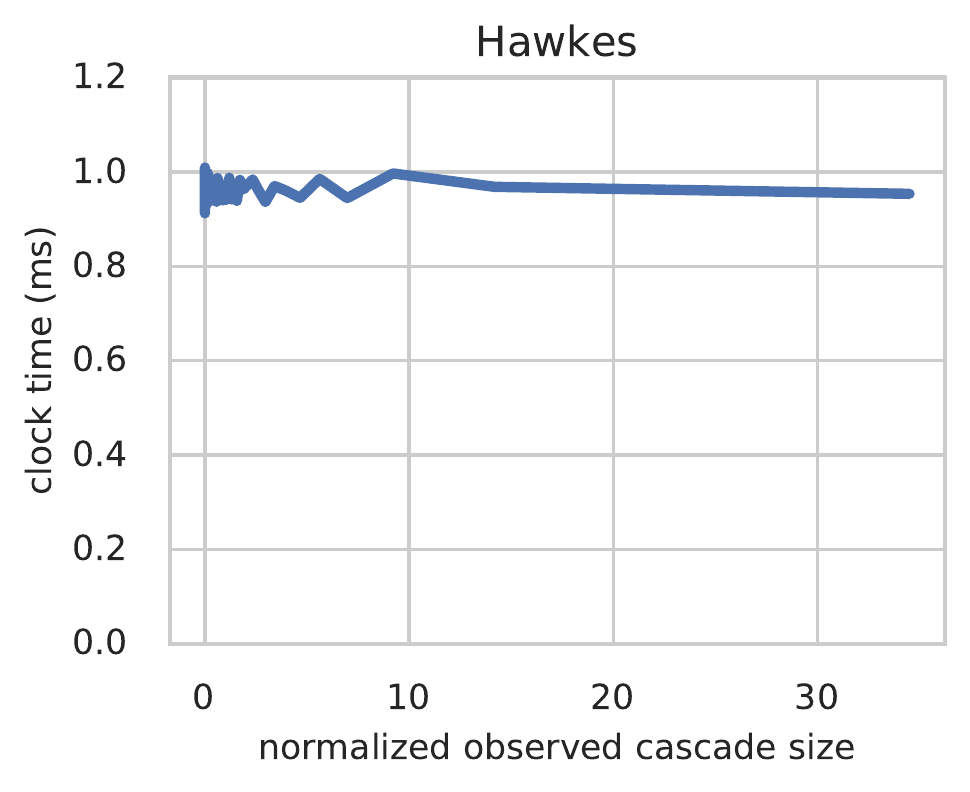}
		\includegraphics[width=0.49\linewidth]{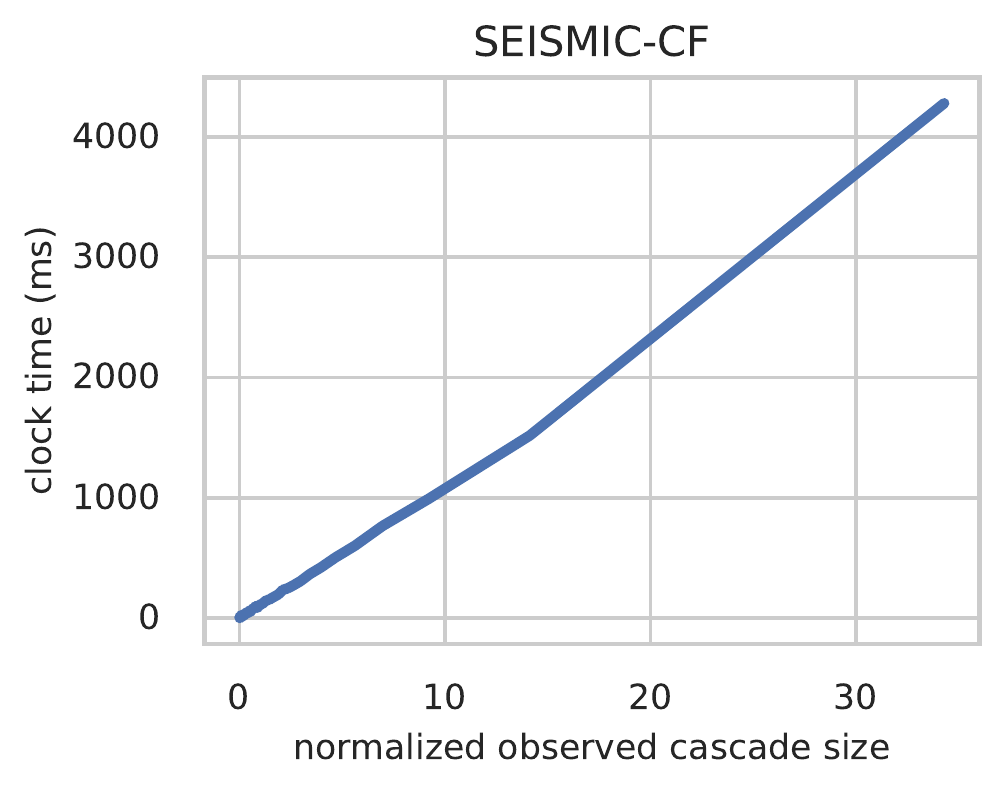}
		\caption{Computation cost of Hawkes and SEISMIC-CF models as a function of the normalised observed cascade size.}
		\label{fig:computation_cost}
	\end{center}
\end{figure}

\subsection{Computation cost of different methods}

\dk{We evaluate computation cost of different methods by measuring the clock time for the computation required to predict the final cascade size on the testing set. We ran these experiments on a sever with 24 Intel Core Processor (Broadwell) CPUs and 114GB of RAM. In Figure~\ref{fig:computation_cost} we report the mean clock time in milliseconds for generating predictions on cascades of different observed sizes (normalized by the average value) in SEISMIC-CF and Hawkes models.}

\dk{As anticipated in Section~\ref{sec:complexity}, the computational cost for SEISMIC-CF scales \mvrev{linearly with} the observed cascade size $N(s)$. Indeed, it can vary 4000x between predictions on cascades with a handful of observed events and cascades with millions of observed events. This is because SEISMIC-CF model requires a pass through all events in the history of the cascade to yield a prediction. As discussed in Sec~\ref{sec:complexity}, other considered models require multiple passes through the observed history of a cascade to produce a prediction for each content item and hence their computation complexity will increase even faster.} 
	
\dk{In contrast, our proposed Hawkes model has a constant computation time for making predictions of any observed cascade size. This is because it only requires an inference from few gradient boosted decision tree models. The static and temporal features we use in the model (discussed in detailes in Appendix~\ref{app:features}) can be computed efficiently at prediction time. For instance, the temporal features in our model constitute simple counters of events in the observed history of a cascade. These counters can be tracked efficiently with a dedicated data structure and fetched in constant time with respect to the cascade history size ~\cite{datar2002maintaining}.}
	
\dk{This result confirms our theoretical findings and suggests that our proposed model can effectively operate at Facebook scale. }

\section{Conclusion}
\label{sec:conc}

We proposed \mvrev{a model} for popularity prediction of social media items that \mvrev{satisfies} a set of design \feedback{considerations} that arise in large-scale online platforms. These \feedback{considerations} include providing accurate predictions for any given prediction \mvrev{time and horizon}, having a constant-time computation complexity at prediction time, and leveraging both static and temporal features to ensure accurate predictions. The model requires combining only a few point predictors, including prediction of the view count acquired up to one or more fixed reference time horizons and a predictor of the effective popularity growth rate. The prediction accuracy is shown to be competitive to separately trained models for specific prediction time horizons, using a large collection of post sharing on Facebook. 

Future work may further explore the space of scalable popularity prediction methods, and study the trade-off between computation complexity and prediction accuracy.

\bibliographystyle{ACM-Reference-Format}
\bibliography{references}

\clearpage

\appendix

\section{Appendix}

\subsection{Example cascades}
\label{app:examples}

\begin{figure}[b!]
	\centering
	\includegraphics[width=\linewidth]{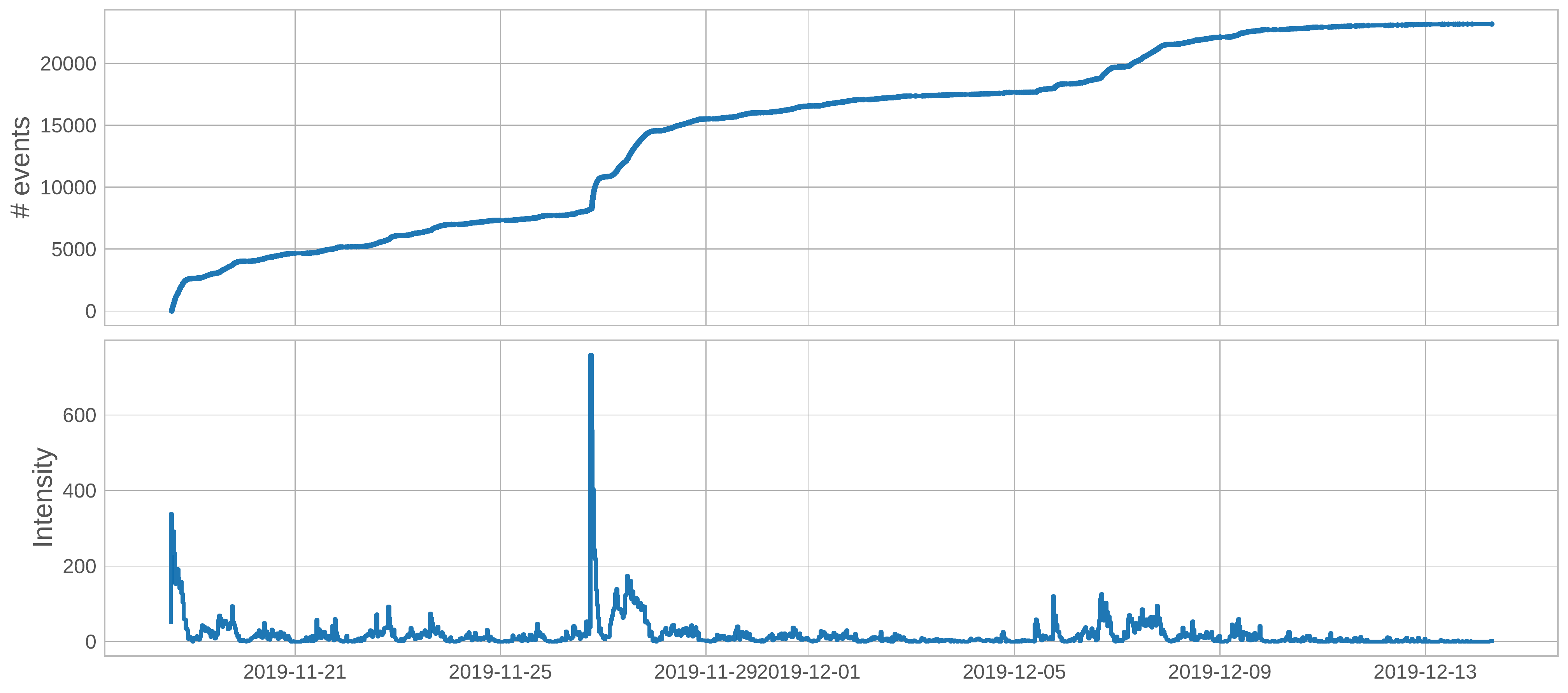}
	\caption{Example Facebook page post: cumulative number of views (top) and views per 30-min time interval (bottom).
	}
	\label{fig:intensity_example}
\end{figure}

\begin{figure}[b!]
	\centering
	\includegraphics[width=\linewidth]{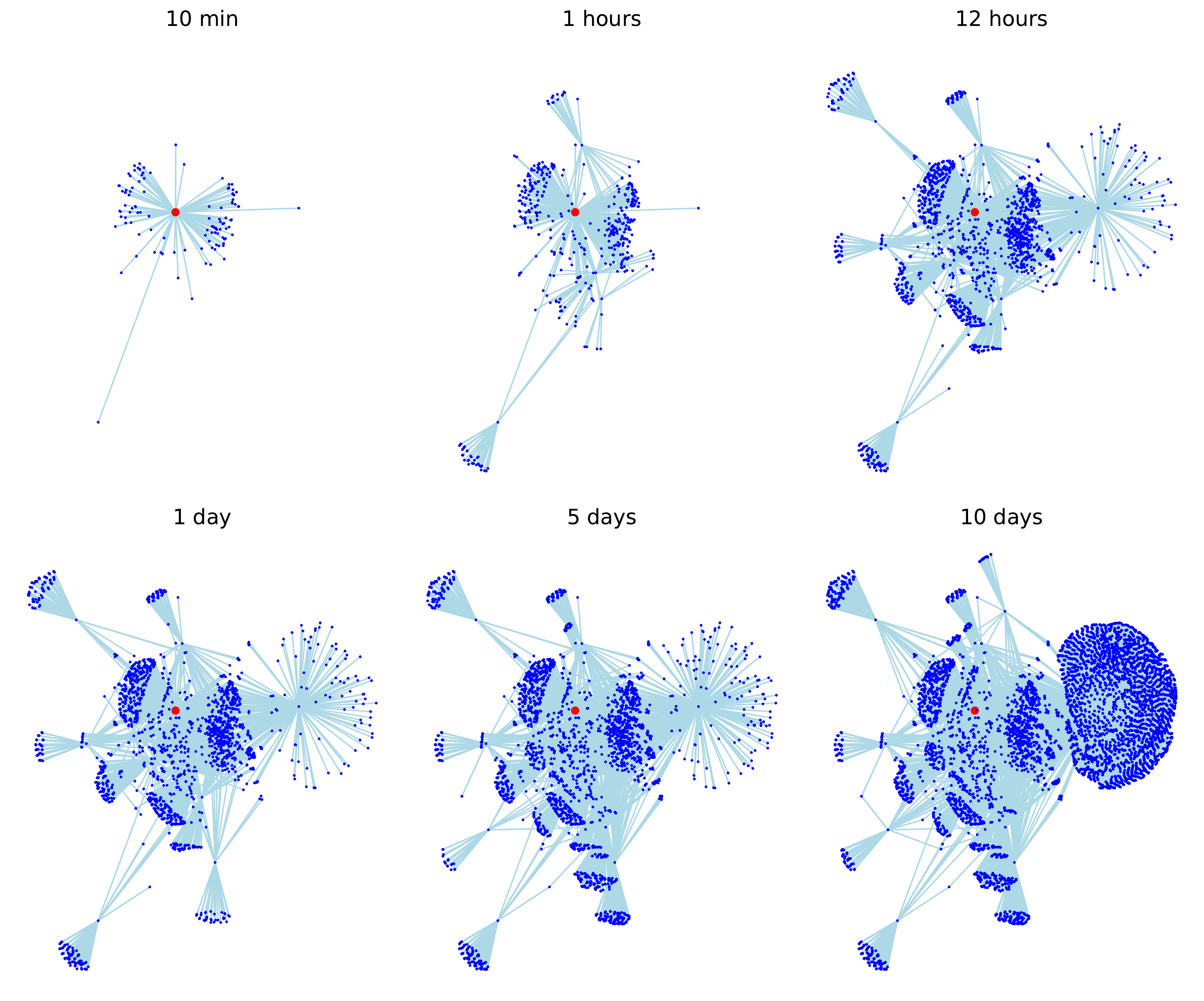}     
	\caption{Information diffusion graph for our example post over time. Each node denotes a user, each edge~--- interaction through the example in Figure \ref{fig:intensity_example}.}
	\label{fig:diffusion_graph}
\end{figure}

Figure~\ref{fig:intensity_example} provides the popularity growth of a single Facebook page post, which has several bursts of view activity, some occurring soon after content creation and some occurring a few days later. 
This example involves substantial content re-sharing, resulting in content re-sharing graphs shown in Figure~\ref{fig:diffusion_graph}. Content views are accumulated by users viewing the content item directly from the post of the content author or indirectly through a chain of re-share posts. Content re-sharing events and post privacy settings govern the information spread in the network, with each re-share providing access to information to some uninformed users. We may think of content view counts as of a superposition of view counts triggered by content re-sharing events.

\begin{figure}[b!]
 	\centering
 	\includegraphics[width=\linewidth]{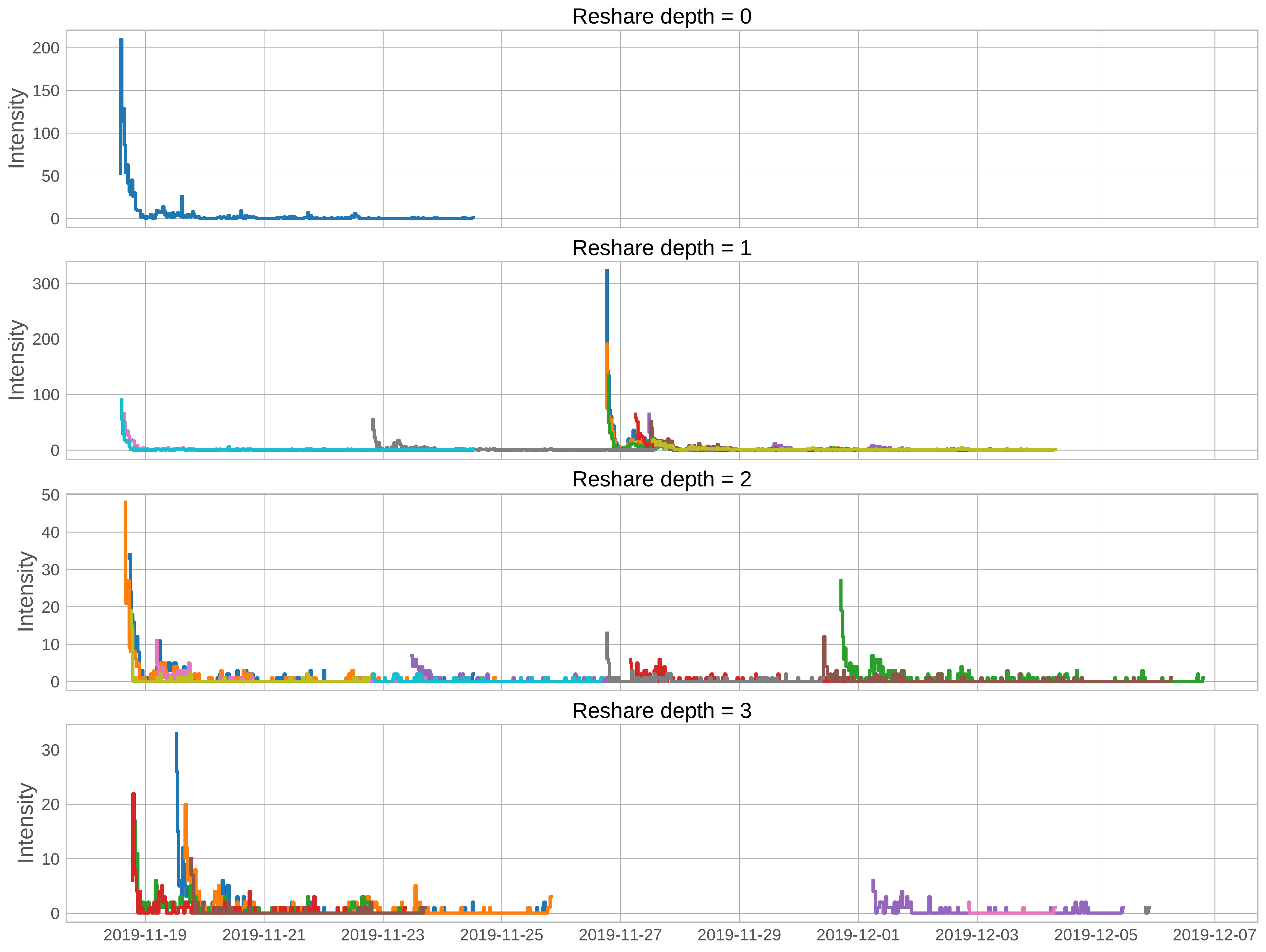}     
 	\caption{Intensity of content view events at different reshare depths of the information diffusion in Figure \ref{fig:intensity_example}.
 	}
 	\label{fig:intensity_breakdown}
 \end{figure}

 In Figure~\ref{fig:intensity_breakdown}, we show a breakdown of content view events by conditioning on the source of information (either the author of original post or the user who re-shared the original post) at different re-share depths (hop distance to the original information source). We can observe how content view counts induced by re-share events result in inflection points in the aggregate cumulative content view counts.

 \subsection{Properties of cascades}

\begin{figure}[b!]
	\centering
	\includegraphics[width=0.48\linewidth]{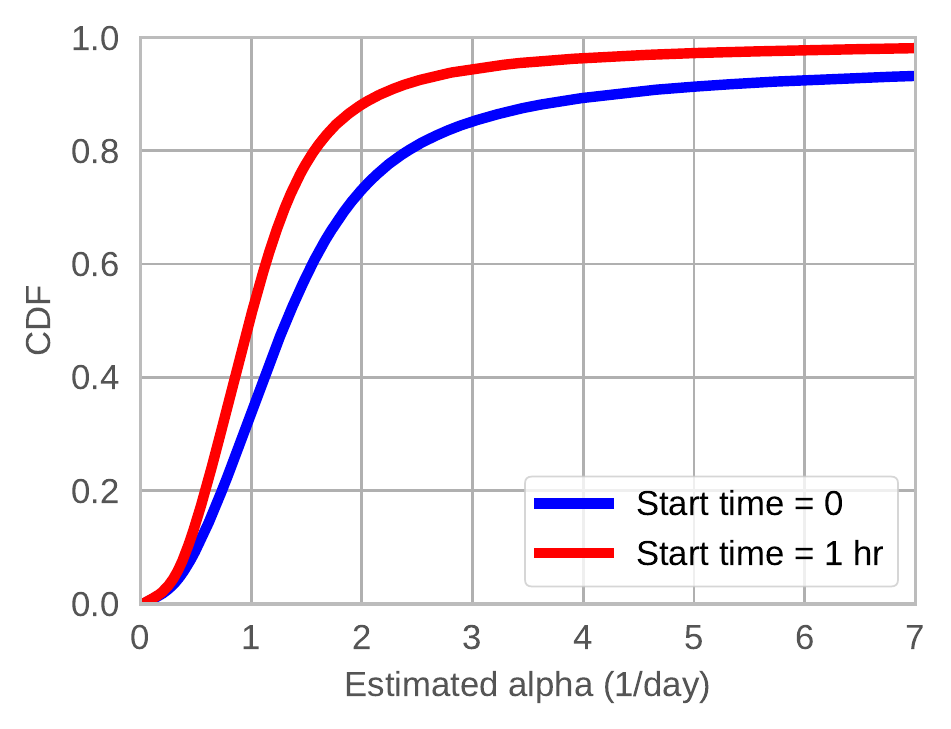}
	\includegraphics[width=0.48\linewidth]{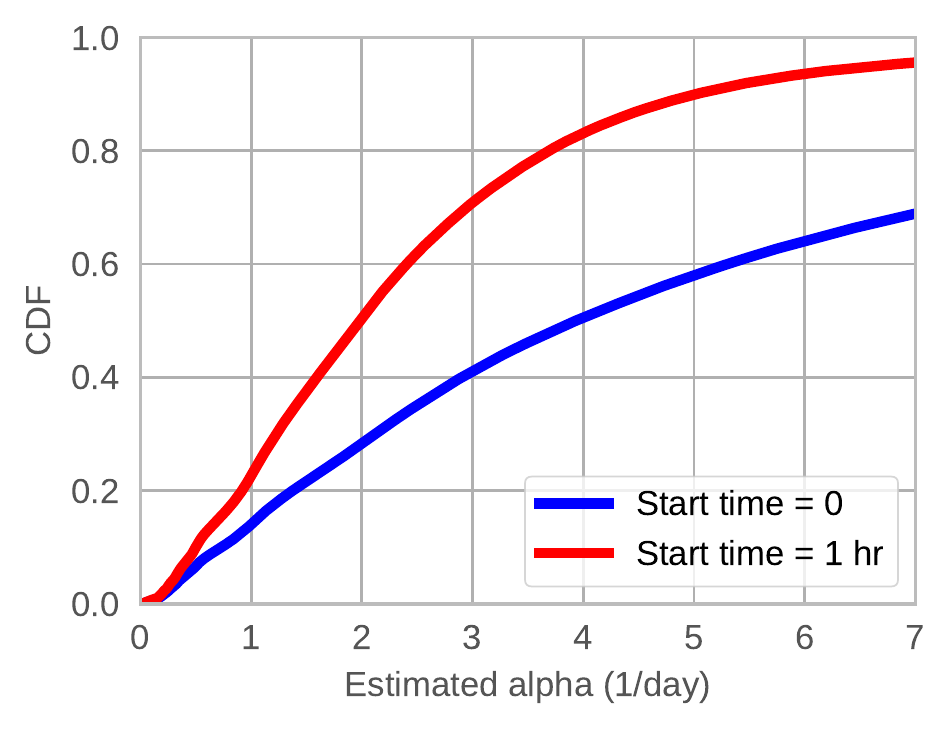}
	\caption{Distribution of effective growth parameter estimates: mean value (left) and median value (right) estimator.}
	\label{fig:alpha_mean_cdf}
\end{figure}

\paragraph{Cascade size and duration} A basic property of a content view count function is the total number of views accumulated over a large time horizon~--- \emph{cascade size}. Another one is \emph{cascade duration}, which characterizes the timeframe within which a piece of content keeps accumulating views. As expected, in our dataset we observe both characteristics to have long-tailed distributions. Further, the averaged shape of stochastic intensity functions estimated from the dataset provide empirical evidence that the view event counts follow an exponential-decay trend over horizons spanning multiple days. More details are provided in Appendix~\ref{app:cascades}.

\begin{figure}[b!]
\centering
\includegraphics[width=0.48\linewidth]{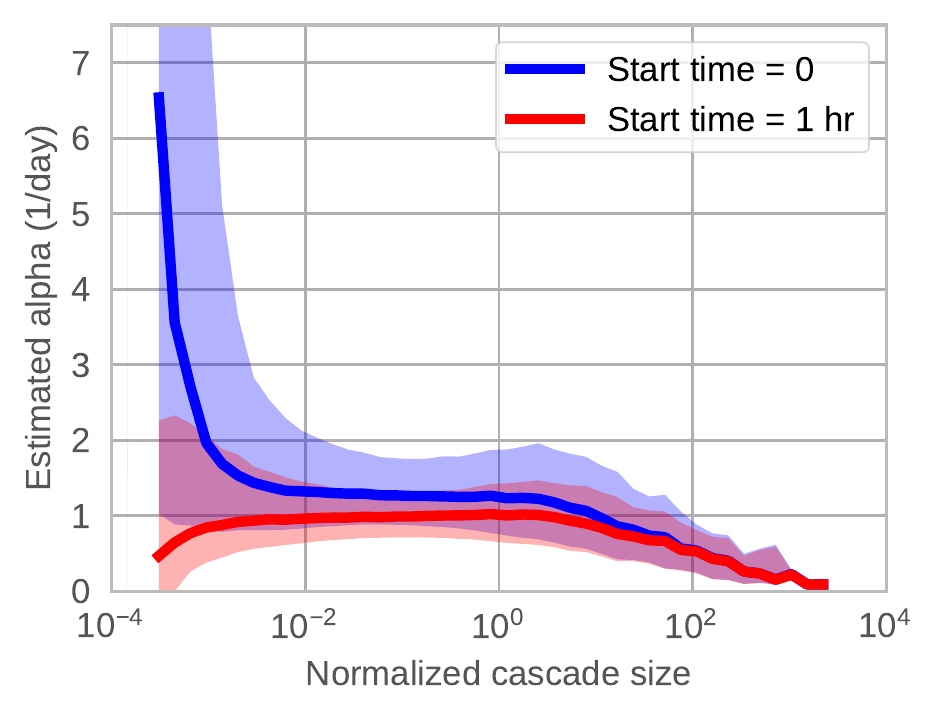} 
\includegraphics[width=0.48\linewidth]{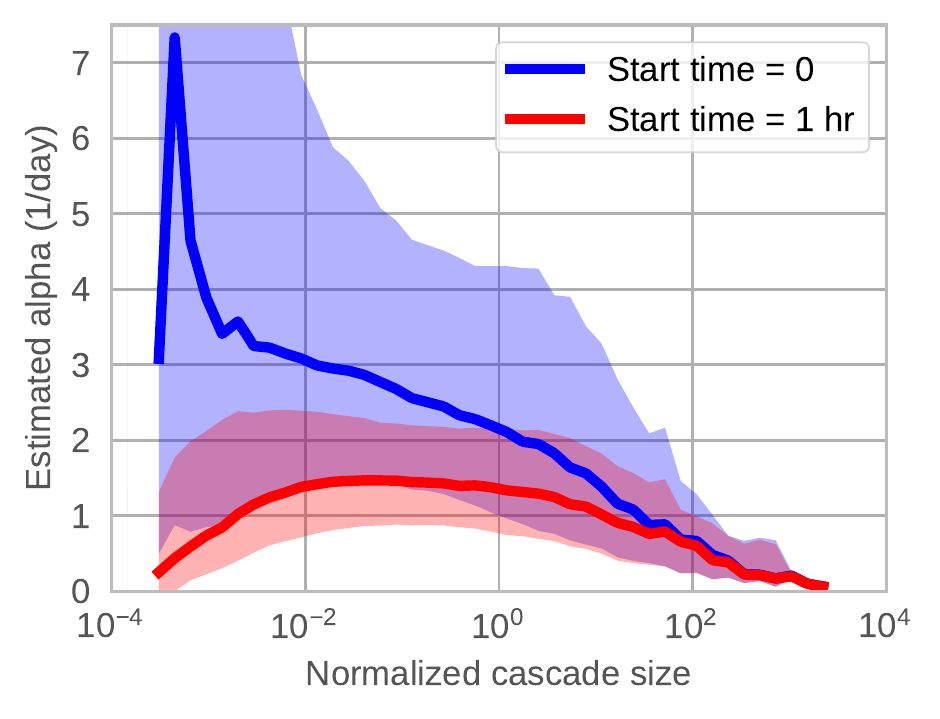} 
\caption{Estimators of the effective growth exponent based on mean (left) and median (right) value versus cascade size.}
\label{fig:alpha_vs_size}
\end{figure}

\paragraph{Effective growth exponent} We examine the mean and quantile value based estimators of the effective growth exponent, defined in Section~\ref{sec:estexp}. Here, we consider the quantile value based estimator with parameter $\gamma = 1/2$, hence we refer to it as a median value based estimator. In Figure~\ref{fig:alpha_mean_cdf}, we show cumulative distribution functions of the estimated effective growth exponents, using either all the event times of a cascade (start time = 0) or only those observed after 1 hour from the content creation time (start time = 1). We observe that the estimated growth exponents cover a wide range of values, with median value of about 1 for the mean value based estimator. The median value based estimator tends to be larger than the mean value based estimator. We attribute this to more weight given to early points by the median value based estimator. The mean value based estimator shows consistent estimates for different values of the time intervals over which the estimate is computed. The median value based estimator shows more discrepancy in this respect and produces larger estimates when excluding an initial time period.  

\paragraph{Effective growth exponent vs. cascade size} We next examine how the effective growth exponent correlates with the cascade size. One may wonder whether the effective growth exponent is largely invariant to the total number of view events accumulated for a content item. In Figure~\ref{fig:alpha_vs_size}, we show the median and quartile values of the estimates conditional on cascade size (normalized by the average value). We observe that the effective growth exponent tends to decrease with cascade size for small cascade sizes but otherwise remains largely invariant. The median value based estimates are more consistent than mean value based estimates when computed by taking only points observed after $1$ hour of the content creation. 

\subsection{Proof of Equation \ref{equ:genlimexp}}
\label{app:genlimexp}

We partition points of a Hawkes point process over different generations with respect to the stochastic intensity components. These generations are defined recursively by defining the $i+1$-st generation points to be those generated by the stochastic intensity kernels associated with the $i$-th generation points. Let $N_i(t)$ be the number of points in $[0,t)$ that belong to the $i$-th generation, for $i \geq 1$. 

Note that for every $s\geq 0$ and $i\geq 1$
$$
\EC{N_{i+1}(+\infty)-N_{i+1}(s)}{ {\mathcal F}_s} = 
\mv{\mu}
\EC{N_{i}(+\infty)-N_{i}(s)}{ {\mathcal F}_s}. 
$$ 
Hence,
$$
\EC{N_{i}(+\infty)-N_{i}(s)}{ {\mathcal F}_s} = 
\mv{\mu^{i-1}}\EC{N_{1}(+\infty)-N_{1}(s)}{ {\mathcal F}_s}.
$$
It follows
\begin{eqnarray*}
\EC{N(+\infty) - N(s) }{ \mathcal{F}_s} &=& \sum_{i=1}^\infty \EC{N_{i}(+\infty)-N_{i}(s)}{ {\mathcal F}_s}\\
&= &  \sum_{i=1}^\infty 
\mv{\mu^{i-1}}\EC{N_{1}(+\infty)-N_{1}(s)}{ {\mathcal F}_s} \\
&=& 
\mv{\frac{1}{1-\mu}}\EC{N_{1}(+\infty)-N_{1}(s)}{ {\mathcal F}_s}.
\end{eqnarray*}
Now, note that
\begin{eqnarray*}
&& \EC{N_{1}(t)-N_{1}(s)}{ {\mathcal F}_s}\\
&=& \int_{s}^t \left(\lambda_0(u) + \sum_{i\geq 1} \phi_{y_i}(u-T_i)\mathbf{1}_{\{0\leq T_i < s\}}\right) du\\
&=& \Lambda_0(t)-\Lambda_0(s) + \sum_{i\geq 1} \left(\Phi_{y_i}(t-T_i)-\Phi_{y_i}(s-T_i)\right)\mathbf{1}_{\{0\leq T_i < s\}}\\
&:=& \Lambda(s,t),
\end{eqnarray*}
where $\Lambda_0$ and $\Phi_y$ are the primitive functions of $\lambda_0$ and $\phi_y$.

Hence, we have
$$
\EC{N(+\infty)-N(s)}{ \mathcal{F}_s} = 
\mv{\frac{1}{1-\mu}}\lim_{t\rightarrow \infty}\Lambda(s,t).
$$

\subsection{Proof of Proposition \ref{prop:exp}}
\label{app:exp}

We consider the Hawkes point process with exponentially decaying intensity, defined by the stochastic intensity in (\ref{equ:phiyexp}). Let $\mu_p := \int_0^\infty y^p dF_Y(y)$. For simplicity of notation, we will write $F$ in lieu of $F_Y$. The process $(\lambda(t), N(t))_{t\geq 0}$ is a continuous-time Markov chain with the infinitesimal generator given by
$$
\mathcal{A} f(\lambda,n) = -\beta \lambda\frac{\partial}{\partial \lambda} f(\lambda,n) + \lambda \left(\int_0^{\infty} f(\lambda+z,n+1)dF(z) - f(\lambda,n)\right).
$$

The following proposition gives the conditional joint Laplace transform and generation function of $(N(t),\lambda(t))$, conditional on the history $\mathcal{F}_s$ observed up to time $t$. Similar characterization is available in Theorem~3.1~\cite{dassios2011} for a more general dynamic contagion process.

\begin{proposition} For any constants $0\leq u\leq 1$, $v\geq 0$ and times $0\leq s\leq t$, the conditional joint Laplace transform and generation function of $(N(t), \lambda(t))$,
\begin{equation}
\psi(u,v) = \EC{u^{N(t)-N(s)} e^{-v \lambda(t)} }{ \mathcal{F}} = e^{-\lambda(s) A(t-s; u,v)},
\label{equ:psijoint}
\end{equation}
where
\begin{equation}
\frac{\partial}{\partial \tau} A(\tau; u,v) = 1 - \beta A(\tau; u,v) - u \psi_F(A(\tau; u,v)),
\label{equ:psi}
\end{equation}
with the boundary condition $A(0; u, v) = v$, and where
\begin{equation}
\psi_F(z) = \int_0^\infty e^{-zx}dF(x).
\label{equ:psiG}
\end{equation}
\end{proposition}

\paragraph{Proof of Proposition~\ref{prop:exp}} From (\ref{equ:psijoint}), we have
\begin{eqnarray}
&& \EC{N(t)-N(s)}{ \mathcal{F}_s} = \lim_{u\uparrow 1,v\downarrow 0}\frac{\partial}{\partial u}\psi(u,v) \nonumber\\
&=& \lambda(s) \lim_{u\uparrow 1, v\downarrow 0} \left(-\left(\frac{\partial}{\partial u}A(t-s; u,v)\right)e^{-\lambda(s)A(t-s; u,v)}\right).\label{equ:fmom}
\end{eqnarray}

From (\ref{equ:psi}) and the boundary condition $A(0; u, v) = v$, we have
\begin{eqnarray}
A(\tau; u,v) &=& v + \tau - \beta \int_0^{\tau} A(x; u,v) dx \nonumber\\
&& - u \int_0^{\tau} \psi_F(A(x; u,v))dx.\label{equ:Aint}
\end{eqnarray}
From this, we have
\begin{eqnarray}
\frac{\partial}{\partial u} A(\tau; u,v) & = & - \beta \int_0^{\tau} \frac{\partial}{\partial u}A(x; u,v)dx - \int_0^{\tau} \psi_F(A(x;u,v))dx\nonumber \\
& & - u\int_0^{\tau}\psi_F'(A(x; u,v))\frac{\partial}{\partial u} A(x;u,v) dx.
\label{equ:long}
\end{eqnarray}
From (\ref{equ:psi}) and the boundary condition $A(0; u,v) = v$, we have
$$
\int_v^{A(\tau; u,v)} \frac{1}{1-\beta x - u \psi_F(x)}dx = \tau.
$$
Since the integrand goes to $\infty$ as $x$ goes to $0$ and $u\uparrow 1$, it follows that 
\begin{equation}
\lim_{u\uparrow 1, v\downarrow 0} A(\tau;u,v) = 0. 
\label{equ:Alim}
\end{equation}
Combining this with (\ref{equ:long}), we have
$$
h(\tau;u) = -(\beta-\mu_1)\int_0^\tau h(x;u) - \tau,
$$
where $h(\tau) : = \lim_{u\uparrow 1, v\downarrow 0} \frac{\partial}{\partial u}A(\tau;u, v)$. Hence, $h(\tau)$ is the solution of the linear ordinary differential equation
$$
\frac{d}{d\tau}h(\tau)  + (\beta - \mu_1) h(\tau) = -1
$$
with initial value $h(0) = 0$. The solution is given by
\begin{equation}
\lim_{u\uparrow 1, v\downarrow 0} \frac{\partial}{\partial u}A(\tau;u, v) = h(\tau) = -\frac{1}{\beta -\mu_1}\left(1-e^{-(\beta-\mu_1)\tau}\right).
\label{equ:partAlim}
\end{equation}
From (\ref{equ:fmom}), (\ref{equ:Alim}) and (\ref{equ:partAlim}), we have
\begin{equation}
\EC{N(t) - N(s)}{ \mathcal{F}_s} = \lambda(s) \frac{1}{\beta -\mu_1}\left(1-e^{-(\beta-\mu_1)(t-s)}\right).
\label{equ:fmom1}
\end{equation}

The asserted expression in the proposition follows by substitution $\mu_1 = \beta \rho_1$. 

\subsection{Proof of Proposition \ref{pro:bounds}}
\label{sec:bounds}

Admit the definitions introduced in Appendix~\ref{app:genlimexp}, and arbitrarily fix the values of the time instances $0\leq s\leq t$. 

The lower bound follows by noting that $\EC{N(t)-N(s)}{\mathcal{F}_s} \geq \EC{N_1(t)-N_1(s)}{\mathcal{F}_s}$ and $\EC{N_1(t)-N_1(s)}{\mathcal{F}_s} = \Lambda(s,t)$. 

To show the upper bound, note that each point has the expected offspring size equal to 
\mv{$\mu$}. Hence, for every $i\geq 1$, we have
$$
\EC{N_{i+1}(t)-N_{i+1}(s)}{\mathcal{F}_s}\leq 
\mv{\mu}\EC{N_i(t)-N_i(s)}{\mathcal{F}_s}.
$$
From this, it follows
\begin{eqnarray*}
\EC{N(t)-N(s)}{\mathcal{F}_s} &=& \sum_{i=1}^\infty \EC{N_i(t)-N_i(s)}{\mathcal{F}_s}\\
&\leq& \sum_{i=1}^\infty 
\mv{\mu^{i-1}}\EC{N_1(t)-N_1(s)}{\mathcal{F}_s}\\
&=& 
\mv{\frac{1}{1-\mu}}\Lambda(s,t).
\end{eqnarray*}

\subsection{Conditional variance}
\label{sec:expvar}

The Hawkes point processes with exponentially decaying intensity allow us also to explicitly characterize higher-order conditional moments. We next present an explicit characterization of the conditional variance of the number of points over an arbitrary time horizon, given the history of the point process up to a time instance. This quantity is of interest to assess the prediction error due to stochasticity of the point process. 

\begin{proposition} For the Hawkes point process with exponentially decaying intensity, for every $0 \leq s \leq t$, we have
\begin{eqnarray}
\VC{N(t) - N(s) }{ \mathcal{F}_s}  &=& \frac{\lambda(s)}{\alpha} \left( \beta^2 \rho_2(1-e^{-2\alpha(t-s)})\right .\nonumber\\
&& \left . + (1-2\beta\rho_1)(1-e^{-\alpha(t-s)}) \right .\label{equ:var}\nonumber\\
&&  \left . + 2(\beta^2\rho_2-\beta\rho_1) \alpha(t-s) e^{-\alpha(t-s)} \right).\nonumber
\end{eqnarray}
\label{prop:expvar}
\end{proposition}

Note that for every fixed $s$, the limit value of the conditional variance as $t$ goes to infinity is equal to
\begin{equation}
\lim_{t\rightarrow \infty}\VC{N(t) - N(s) }{ \mathcal{F}_s} = \Sigma^2 \frac{1}{\alpha}\lambda(s)
\label{equ:limvar}
\end{equation}
where 
\begin{equation}
\Sigma^2 = (1-\beta \rho_1)^2 + \beta^2 \sigma^2.
\label{equ:Sigma}
\end{equation}

From (\ref{equ:psijoint}), we have
\begin{eqnarray}
&& \EC{(N(t)-N(s))^2}{ \mathcal{F}_s} - \EC{N(t)-N(s)}{ \mathcal{F}_s}\nonumber \\
& = & \lim_{u\uparrow 1, v\downarrow 0}\frac{\partial^2}{\partial u^2} \psi(u,v).
\label{equ:smom}
\end{eqnarray}
Now, note
\begin{eqnarray}
\frac{\partial^2}{\partial u^2} \psi(u,v) &=& \lambda(s) \left(\lambda(s)\left(\frac{\partial}{\partial u} A(t-s;u,v)\right)^2\right .\nonumber\\ 
&& \left .-\frac{\partial^2}{\partial u^2}A(t-s;u,v)\right)e^{-\lambda(s)A(t-s;u,v)}.
\label{equ:partpsi2}
\end{eqnarray}

By (\ref{equ:partAlim}), we have
\begin{eqnarray}
&& \lim_{u\uparrow 1, v\downarrow 0}\lambda(s)\left(\frac{\partial}{\partial u} A(t-s;u,v)\right)^2\nonumber\\ 
&=& \lambda(s)\frac{1}{(\beta-\mu_1)^2}\left(1-e^{-(\beta-\mu_1)(t-s)}\right)^2.
\label{equ:fterm}
\end{eqnarray}
It remains to evaluate the term $\lim_{u\uparrow 1, v\downarrow 0} \frac{\partial^2}{\partial u^2} A(t-s; u,v)$.

From (\ref{equ:Aint}), we obtain
\begin{eqnarray*}
\frac{\partial^2}{\partial u^2} A(\tau; u,v) &=& - \beta \int_0^\tau \frac{\partial^2}{\partial u^2}A(x; u,v)dx\\
&& - 2 \int_0^\tau \psi_F'(A(x;u,v))\frac{\partial}{\partial u}A(x;u,v) dx \\
&& - u\int_0^\tau \psi_F''(A(x;u,v))\left(\frac{\partial}{\partial u}A(x;u,v)\right)^2dx\\ 
&& - u \int_0^\tau \psi_F'(A(x;u,v))\frac{\partial^2}{\partial u^2} A(x;u,v)dx. 
\end{eqnarray*}

Now, using the facts:
\begin{eqnarray*}
\lim_{u\uparrow 1, v\downarrow 0} A(x;u,v) &=& 0\\
\lim_{z\rightarrow 0} \psi_F'(z) &=& -\mu_1\\
\lim_{z\rightarrow 0} \psi_F''(z) &=& \mu_2
\end{eqnarray*}
and letting $g(\tau) := \lim_{u\uparrow 1, v\downarrow 0}\frac{\partial^2}{\partial u^2} A(\tau;u,v)$, we have
$$
g(\tau) = -(\beta-\mu_1) \int_0^\tau g(x) dx + 2\mu_1\int_0^\tau h(x)dx - \mu_2 \int_0^\tau h(x)^2 dx,
$$
where recall $h(x)$ is given by (\ref{equ:partAlim}). Hence, $g(\tau)$ is the solution of the linear ordinary differential equation
$$
\frac{d}{d\tau} g(\tau) + (\beta-\mu_1) g(\tau) = 2\mu_1h(x)-\mu_2 h(\tau)^2
$$
with initial value $g(0) = 0$. The solution is
\begin{eqnarray*}
g(\tau) &=& e^{-(\beta-\mu_1)\tau} \int_0^\tau e^{(\beta-\mu_1)x}[2\mu_1 h(x)-\mu_2 h(x)^2]dx\\
&=& e^{-(\beta-\mu_1)\tau}\left[2\mu_1\int_0^\tau (e^{(\beta-\mu_1)x}-1)dx \right .\\
&& \left . - \mu_2 \int_0^\tau (e^{(\beta-\mu_1)x} -2 + e^{-(\beta-\mu_1)x})dx\right]\\
&=& e^{-(\beta-\mu_1)\tau}\left[\frac{2\mu_1}{\beta-\mu_1}(e^{(\beta-\mu_1)\tau}-1) - 2\mu_1 \tau \right .\\
&& \left .  - \mu_2\left(\frac{1}{\beta-\mu_1}(e^{(\beta-\mu_1)\tau}-1) - 2\tau + \frac{1}{\beta-\mu_1}(1-e^{-(\beta-\mu_1)\tau})\right)\right]\\
&=& \frac{2\mu_1}{\beta-\mu_1}(1-e^{-(\beta-\mu_1)\tau}) - 2\mu_1 \tau e^{-(\beta-\mu_1)\tau} \\
&& - \mu_2\left(\frac{1}{\beta-\mu_1}(1-e^{-(\beta-\mu_1)\tau}) - 2\tau e^{-(\beta-\mu_1)\tau} \right .\\
&& \left . + \frac{1}{\beta-\mu_1}e^{-(\beta-\mu_1)\tau}(1-e^{-(\beta-\mu_1)\tau})\right)\\
&=& \frac{2\mu_1}{\beta-\mu_1}(1-e^{-(\beta-\mu_1)\tau}) - 2(\mu_1-\mu_2) \tau e^{-(\beta-\mu_1)\tau} \\
&& - \frac{\mu_2}{\beta-\mu_1}(1-e^{-2(\beta-\mu_1)\tau})
\end{eqnarray*}

Combining this with (\ref{equ:fterm}) and (\ref{equ:partpsi2}), we obtain
\begin{eqnarray*}
\lim_{u\uparrow 1, v\downarrow} \frac{\partial^2}{\partial u^2} \psi(u,v) &=& \lambda(s)^2\frac{1}{(\beta-\mu_1)^2}(1-e^{-(\beta-\mu_1)(t-s)})^2\\
&& + \lambda(s)\frac{\mu_2}{\beta-\mu_1}(1-e^{-2(\beta-\mu_1)(t-s)})\\
&&  - \lambda(s)\frac{2\mu_1}{\beta-\mu_1}(1-e^{-(\beta-\mu_1)(t-s)})\\
&& + \lambda(s) 2(\mu_2-\mu_1) (t-s) e^{-(\beta-\mu_1)(t-s)}.
\end{eqnarray*}

Using this in (\ref{equ:smom}), we have
\begin{eqnarray*}
\EC{(N(t)-N(s))^2 }{ \mathcal{F}_s} &=& \lambda(s)^2\frac{1}{(\beta-\mu_1)^2}(1-e^{-(\beta-\mu_1)(t-s)})^2\\
&& + \lambda(s)\frac{\mu_2}{\beta-\mu_1}(1-e^{-2(\beta-\mu_1)(t-s)})\\
&&  + \lambda(s)\frac{1-2\mu_1}{\beta-\mu_1}(1-e^{-(\beta-\mu_1)(t-s)})\\
&& + \lambda(s) 2(\mu_2-\mu_1) (t-s) e^{-(\beta-\mu_1)(t-s)}. 
\end{eqnarray*}

From this and (\ref{equ:fmom1}), we have that the conditional variance of $N(t) - N(s)$ is given by
\begin{eqnarray*}
\VC{N(t)-N(s)}{ \mathcal{F}_s} &=& \lambda(s)\left( \frac{\mu_2}{\beta-\mu_1}(1-e^{-2(\beta-\mu_1)(t-s)})\right .\\
&&  + \frac{1-2\mu_1}{\beta-\mu_1}(1-e^{-(\beta-\mu_1)(t-s)}) \\
&&\left . + 2(\mu_2-\mu_1) (t-s) e^{-(\beta-\mu_1)(t-s)} \right).
\end{eqnarray*}

The asserted expression in the proposition follows by the substitution $\mu_1 = \beta \rho_1$ and $\mu_2 = \beta^2 \rho_2$.

\subsection{Variance of the cascade size}
\label{sec:variance}

We consider the coefficient of variation of $N(t)$ for asymptotically large $t$, conditional on the history $\mathcal{F}_s$ observed up to time $s$, which is given by
$$
\lim_{t\rightarrow \infty}\frac{\sqrt{\mathrm{Var}[N(t)\mid \mathcal{F}_s]}}
{\EC{N(t)}{\mathcal{F}_s}} 
= \Sigma \sqrt{
\frac{1}{\EC{N(+\infty)}{ \mathcal{F}_s}}
\left(1-\frac{N(s)}{\EC{N(+\infty)}{\mathcal{F}_s}}\right)
}.
$$
In particular, for $s = 0$ and $N(s) = 0$, we have
$$
\lim_{t\rightarrow \infty}\frac{\sqrt{\VC{N(t)}{ \lambda(0)}}}
{\EC{N(t)}{\lambda(0)}} = \Sigma\frac{1}{\sqrt{\EC{N(+\infty)}{\lambda(0)}}}.
$$

If we take $\EC{N(+\infty)}{\lambda(0)} = \lambda(0)/\alpha = n$, were $n$ is a scaling parameter, we have 
$$
\lim_{t\rightarrow \infty}\frac{\sqrt{\VC{N(t)}{ \lambda(0)}}}
{\EC{N(t)}{\lambda(0)}} = \Sigma\frac{1}{\sqrt{n}}.
$$

\begin{figure}
\begin{center}
\includegraphics[width=0.15\textwidth]{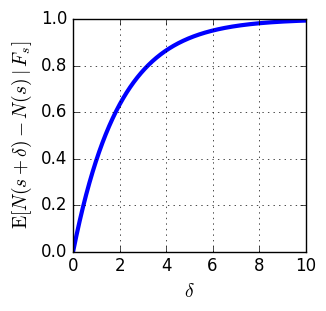}
\includegraphics[width=0.15\textwidth]{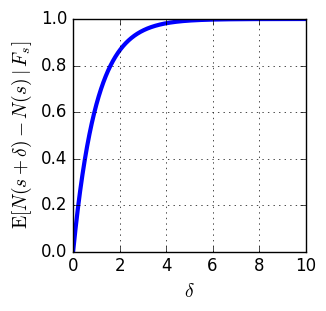}
\includegraphics[width=0.15\textwidth]{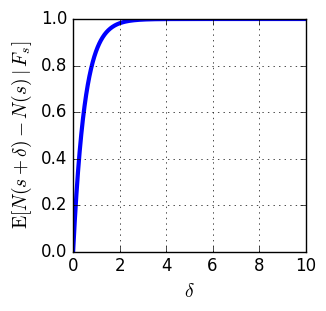}\\
\includegraphics[width=0.15\textwidth]{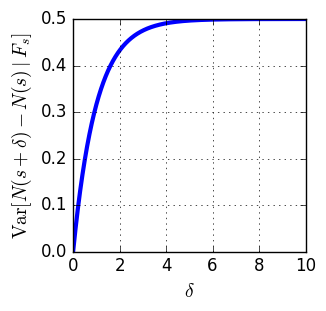}
\includegraphics[width=0.15\textwidth]{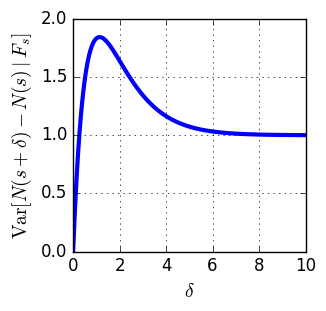}
\includegraphics[width=0.15\textwidth]{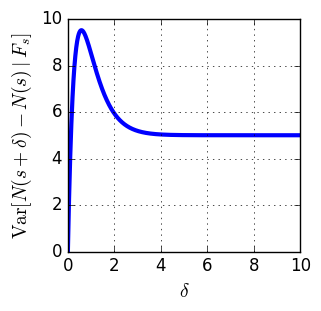}
\caption{(Top) conditional expected value of the count increment and (bottom) conditional variance of the count increment for $\lambda(s)/\alpha = 1$ and $\beta = 1, 2, 4$ from left to right.}
\label{fig:ev}
\end{center}
\end{figure}

\subsection{Simple numerical example} 

In Figure~\ref{fig:ev} we illustrate how the conditional expected value and variance of the count depend on time. Notably, the conditional variance peaks at a certain time instance and converges to a limit whose value is characterized in Eq. (\ref{equ:limvar}).   

\subsection{Mean value based estimator of the effective growth exponent}
\label{sec:meanalpha}

We first prove the following equation
$$
\EC{\int_s^\infty (N(+\infty)-N(t))dt }{\mathcal{F}_s} = \frac{1}{\alpha} \EC{N(+\infty)-N(s)}{\mathcal{F}_s}. 
$$

For any $s\geq 0$, we have
\begin{eqnarray*}
&& \EC{\int_s^\infty (N(+\infty)-N(t))dt }{\mathcal{F}_s}\\
&=& \EC{\int_s^\infty \EC{N(+\infty)-N(t)}{\mathcal{F}_t} dt }{ \mathcal{F}_s} \\
&=& \frac{1}{\alpha}\EC{\int_s^\infty \lambda(t)dt}{\mathcal{F}_s}\\
&=& \frac{1}{\alpha} \EC{N(+\infty)-N(s)}{\mathcal{F}_s}. 
\end{eqnarray*}

We next show that
$$
\int_0^\infty (n-N(t))dt = \sum_{i=1}^n T_i
$$
which follows by simple calculus
\begin{eqnarray*}
\int_0^\infty (n-N(t))dt &=& \sum_{i=0}^{n-1} \int_{T_i}^{T_{i+1}}(n-i) dt\\
&=& \sum_{i=0}^{n-1}(T_{i+1}-T_i)(n-i)\\
&=& \sum_{i=1}^n T_i.
\end{eqnarray*}

\subsection{Quantile value based estimator of the effective growth exponent}
\label{sec:quantileerror}

In this section, we provide a bound for the bias of the quantile value based estimator of the effective growth exponent. In particular, we will show that $\E[\hat{\alpha}]\geq \Omega(1/\log(n))\alpha$, for the quantile value based estimator when $\lambda(0)=\alpha n$ and  $\gamma = 1-1/n$, where $n$ is a scaling parameter. 

Let us define 
$$
f_\gamma(a) = \EC{T_\gamma }{ \lambda(0) = a}.
$$

\begin{proposition} Function $f_\gamma$ satisfies the following inequality, for every $a\geq 0$,
\begin{eqnarray*}
f_\gamma(a) & \leq & \frac{1}{\alpha}\left(\log\left(\frac{1}{1-\gamma}\right) + \gamma \EC{\frac{\lambda(\tau_\gamma)}{\alpha N(\tau_\gamma)} \mathbf{1}_{\{N(\tau_\gamma) > 0\}} }{ \lambda(0)=a}\right)\\
&& + f_\gamma\left(a\left(1-\gamma\right)^{\frac{1}{1-\rho_1}}\right)\PC{N(\tau_\gamma) = 0 }{ \lambda(0) = a}.
\end{eqnarray*}
\label{prop:etg}
\end{proposition}

From this proposition, we have the following corollary:

\begin{corollary} For any fixed $\beta > 0$, $0 \leq \rho_1 < 1$, and initial intensity set such that $\lambda(0) = \alpha n$, by taking $\gamma = 1-1/n$, we have
$$
\EC{T_{1-1/n} }{ \lambda(0) = \alpha n } \leq \frac{1}{\alpha}(\log(n) + 1 + o(1)).
$$
\label{prop:etg1}
\end{corollary}

The corollary implies the following estimation guarantee for the effective growth exponent $\alpha$: for $\lambda(0) = \alpha n$ and $\gamma = 1 - 1/n$,
$$
\E[\hat{\alpha}]\geq \frac{1}{\EC{T_\gamma }{ \lambda(0)}} \geq \frac{1-o(1)}{\log(n)+1}\alpha.
$$

\subsection{Predicting relative growth}
\label{app:relative}
In this section, we present how our framework can be extended for prediction of relative cascade growth, where the goal is to predict whether the cascade size will eventually exceed a given factor of its current size. A special instance of this problem was considered in \cite{CADKL14} asking to predict whether a cascade will double in size.

Our relative growth prediction problem can be formulated as follows: given the observed history $\mathcal{F}_s$ at time $s$ and parameter $c > 1$, the goal is to predict whether the count $N(t)$ will eventually be larger or equal than $c N(s)$. Assume that points are according to a Hawkes point process with exponentially decaying intensity with the effective growth exponent $\alpha$.

Using (\ref{equ:expn}), we note that $\E[N(+\infty)\mid \mathcal{F}_s] \geq c N(s)$ is equivalent to
\begin{equation}
\lambda(s) \geq (c-1)\alpha N(s).
\label{equ:thre}
\end{equation}
Intuitively, the condition requires the stochastic intensity to be larger than a threshold that is proportional to the current count value. The following proposition gives a condition that accounts for stochasticity of the point process.

\begin{proposition} For any constant $0 < \delta \leq 1$, given history $\mathcal{F}_s$ at time $s$ and constant $c > 1$ such that $\lambda(s) > (c-1)\alpha N(s)$, we have $N(+\infty) > c N(s)$ with probability at least $1-\delta$, if
\begin{equation}
\lambda(s) \geq \left(c-1 + \chi(N(s))\right)\alpha N(s),
\label{equ:thre1}
\end{equation}
where 
$$
\chi(x) := \frac{\Sigma^2}{2\delta x} + \sqrt{2 (c-1) \frac{\Sigma^2}{2\delta x} + \left(\frac{\Sigma^2}{2\delta x}\right)^2}
$$
and $\Sigma$ is defined in (\ref{equ:Sigma}).
\label{pro:rg}
\end{proposition}

The proposition tells us that a simple threshold decision rule can be used, similar to (\ref{equ:thre}), but with a threshold that accounts for the variance parameter $\Sigma$ of the point process.

\begin{figure}[t]
	\centering
	\includegraphics[width=0.48\linewidth]{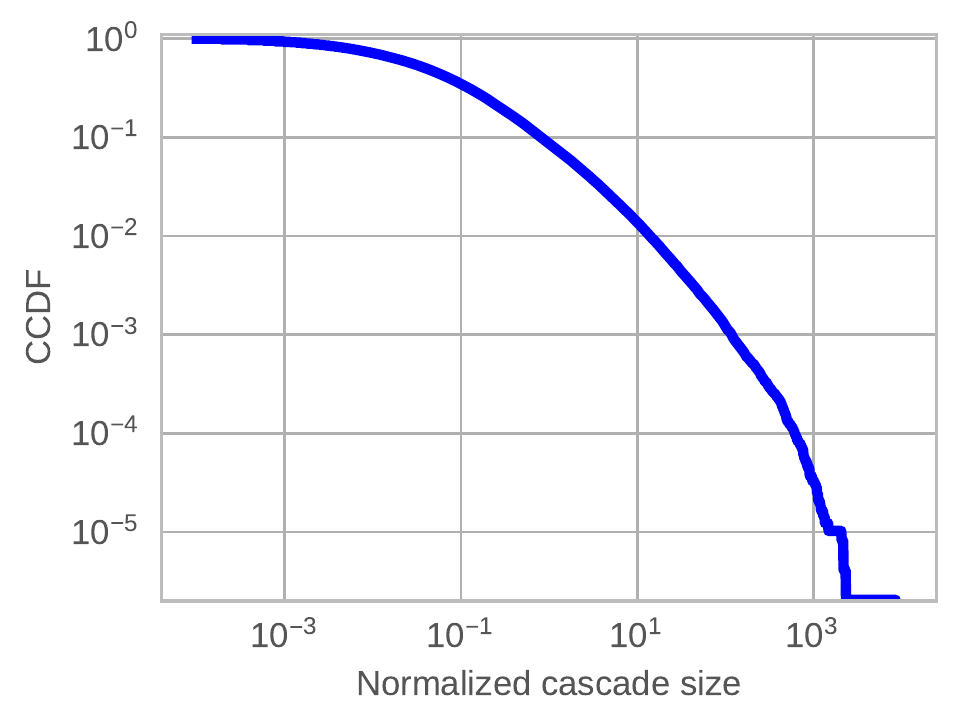}\hspace*{2mm}
	\includegraphics[width=0.48\linewidth]{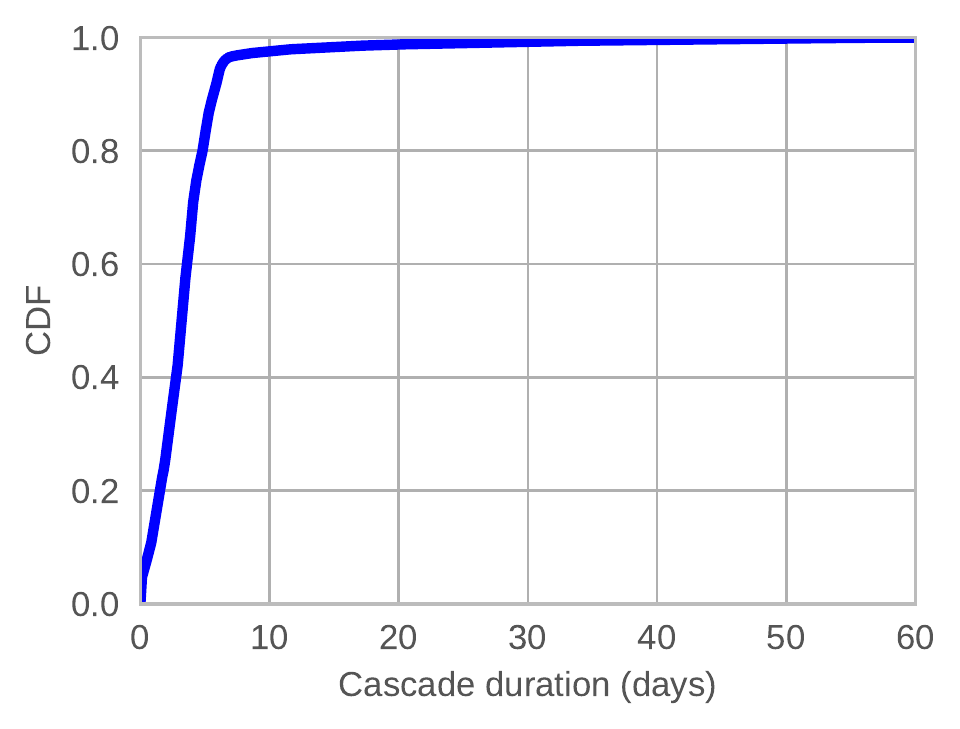}
	\caption{Distribution of cascade size (left) and duration (right).}
	\label{fig:cascade_size}
\end{figure}

\begin{figure}[t]
	\centering
	\includegraphics[width=0.48\linewidth]{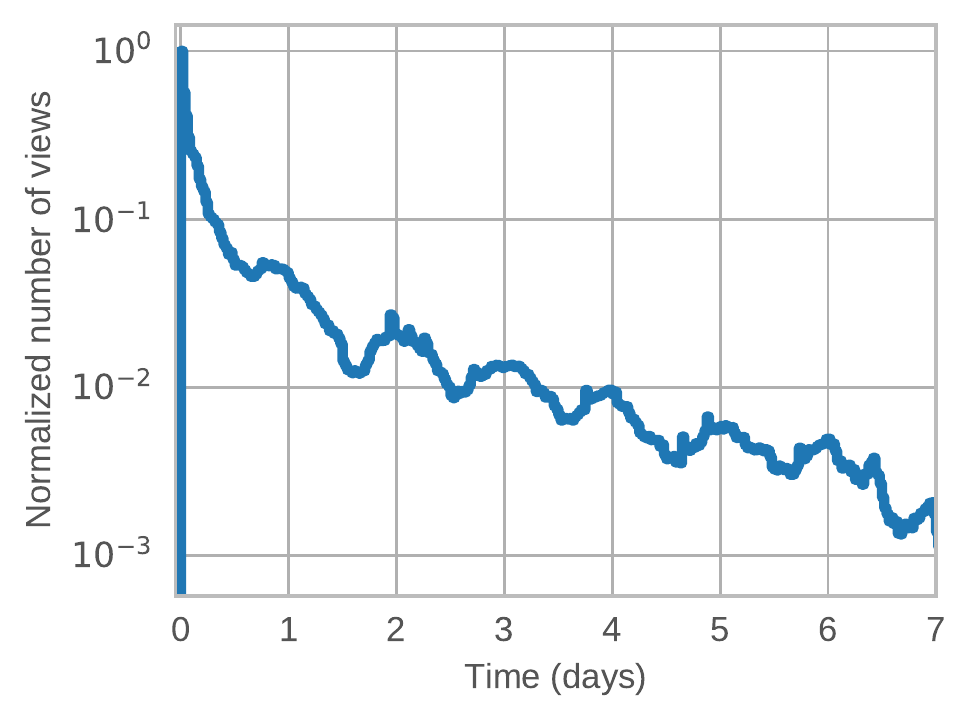}	
	\includegraphics[width=0.48\linewidth]{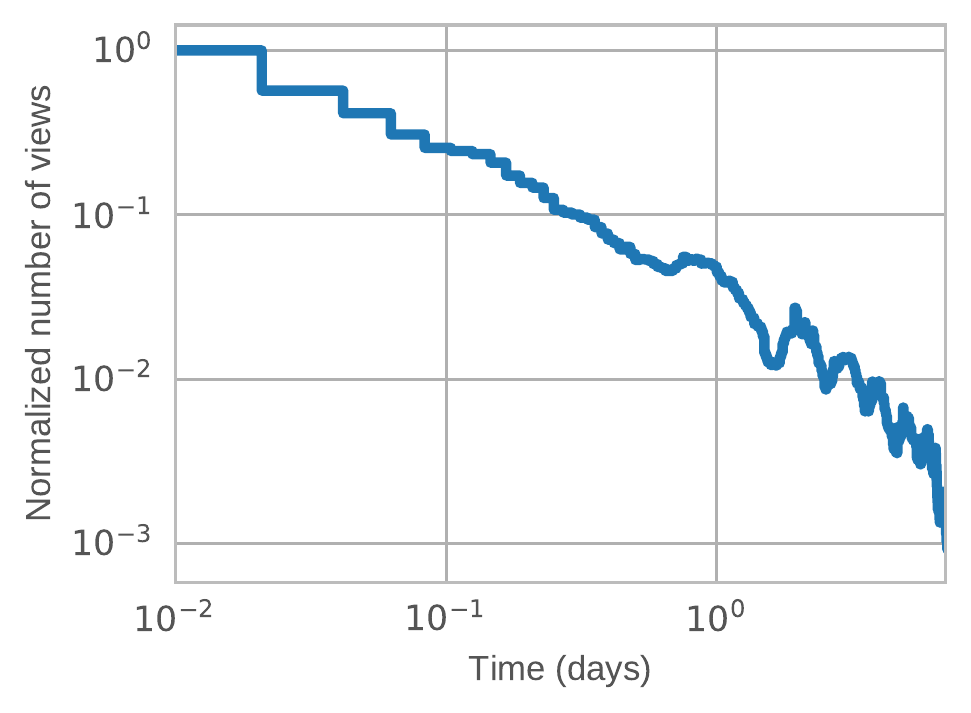}
	\caption{Stochastic intensity vs time.}
	\label{fig:intensity_time}
\end{figure}

\subsection{Properties of cascades in the dataset}
\label{app:cascades}
In Figure~\ref{fig:cascade_size} (left plot), we show complementary cumulative distribution functions of the total number of views observed per content item over the entire observation interval (i.e., cascade size), normalized by the average value. 

We define \emph{cascade duration} as the smallest time at which a fixed fraction of the total number of view events of a content item is reached. In our experiments, we set this fraction to be $0.95$. This definition of cascade duration, instead of the maximum time spanned by a cascade is more robust to outliers---content items receiving a small fraction of view events after a long time. In Figure~\ref{fig:cascade_size} (right plot), we observe that most of the content view events are accumulated within one week after the content item creation, with the median value of about 3 days.      

In Figure~\ref{fig:intensity_time}, we show "fresh" view counts over 30 minute time bins aggregated over content items, where the origin corresponds to a content item creation time. This is shown for different scalings of the $x$ and $y$ axis. These graphs exhibit a decreasing trend with local extrema obeying a daily seasonality. Under the hypothesis that counts follow an exponential decrease, we should observe a linear trend for linear $x$ and logarithmic $y$ axes. From Figure~\ref{fig:intensity_time} (left), we observe this to be overall true over a time period spanning several days. Under the hypothesis that counts follow a power-law decrease, we should observe a linear decrease when both $x$ and $y$ axes are in logarithmic scale. From Figure~\ref{fig:intensity_time} (bottom), we observe that the counts do not seem to be consistent with a power-law decay over a time interval spanning several days. 

\subsection{Proof of Proposition~\ref{pro:rg}}

The proof is by a simple application of the Chebyshev's inequality: for any random variable $X$ with expected value $\mu$ and variance $\sigma^2$, $\Pr[|X-\mu|\geq x]\leq \frac{\sigma^2}{x^2}$, for all $x > 0$.

By Chebyshev's inequality and Proposition~\ref{prop:expvar}, we have
\begin{eqnarray}
&& \PC{N(+\infty) \leq c N(s) }{ \mathcal{F}_s} \nonumber\\
&\leq & \Pr\left[\left| N(+\infty) - \EC{N(+\infty)}{ \mathcal{F}_s}\right| \geq \EC{N(+\infty)}{ \mathcal{F}_s}-cN(s)\right] \nonumber\\
&\leq & \frac{\VC{N(+\infty) - N(s) }{ \mathcal{F}_s}}{\left(\EC{N(+\infty)}{\mathcal{F}_s}-cN(s)\right)^2}\nonumber\\
&=& \frac{\frac{\lambda(s)}{\beta(1-\rho_1)}\Sigma^2}{\left(\frac{\lambda(s)}{\beta(1-\rho_1)}- (c-1)N(s)\right)^2 }.\label{equ:cheb}
\end{eqnarray}

Let $a : = \lambda(s)/[\beta(1-\rho_1)]$ and $b = (c-1)N(s)$. Then, requiring that the right-hand side in (\ref{equ:cheb}) is less than or equal to $\delta$ is equivalent to
$$
(a-b)^2 \geq \frac{\Sigma^2}{\delta}a,
$$
which is equivalent to
$$
a^2 - \left(2b + \frac{\Sigma^2}{\delta}\right)a + b^2 \geq 0.
$$
The solution is 
$$
a \geq \frac{2b + \frac{\Sigma^2}{\delta} + \sqrt{\left(2b+\frac{\Sigma^2}{\delta}\right)^2 - 4b^2}}{2}.
$$
Substituting back $a = \lambda(s)/[\beta(1-\rho_1)]$ and $b = (c-1)(N(s)$, after some rearrangements we have
$$
\lambda(s) \geq \left(c-1 + \chi(N(s))\right)\beta(1-\rho_1)N(s),
$$
where
$$
\chi(x) = \frac{\Sigma^2}{2\delta x} + \sqrt{2 (c-1) \frac{\Sigma^2}{2\delta x} + \left(\frac{\Sigma^2}{2\delta x}\right)^2}.
$$
This completes the proof of the proposition.

It is noteworthy that for the Hawkes point process with exponentially decaying intensity, predicting whether the count will eventually exceed factor $c$ of the count at the prediction time $s$ amounts to checking whether the stochastic intensity exceeds a threshold value, which is a function of the count at the prediction time $N(s)$, effective growth exponent $\alpha$, and variance $\Sigma^2$.

\begin{table*}[t]
	\caption{Cardinality and relative importance of different feature categories used for modelling the effective growth exponent $\hat{\alpha}$ and point predictor $Y(\delta^*;s)$.}
	\label{app:tab:features}
	\begin{tabular}{ p{3.5cm}  p{1.2cm}  p{4.3cm}   p{1.5cm}  p{2.4cm}  p{2.6cm} } \hline
		\multicolumn{3}{c}{\textbf{Category of features}}  & \textbf{Number \mbox{of features}} & \textbf{Importance for predicting \mbox{cascade size at $\delta^*$}} & \textbf{Importance for \mbox{predicting} growth exponent $\alpha$} \\ \hline
		\multirow{5}{4cm}{Engagement features} & \multirow{2}{*}{views} 	& on original post 				& 282				& 0.53108 					& 0.25848 \\ \cline{3-6}
		& 	& cumulative on page's other posts 	& 484				& 0.09080 					& 0.31896 \\ \cline{2-6}
		& \multicolumn{2}{l}{shares} 	 						& 276				& 0.03030						& 0.00472 \\ \cline{2-6}
		& \multicolumn{2}{l}{comments}						& 92					& 0.00362						& 0.00033 \\ \cline{2-6}
		& \multicolumn{2}{l}{reactions} 						& 368				& 0.00250						& 0.00001 \\ \cline{2-6}
		& \multicolumn{2}{l}{combinations} 	 					& 5					& 0.07204						& 0.05416 \\ \hline
		\multicolumn{3}{l}{Page features}      						 							& 349 			   	& 0.16319						& 0.32308 \\ \hline	
		\multicolumn{3}{l}{Content features}  						 							& 23 			   		& 0.01100						& 0.01939 \\ \hline
		\multicolumn{3}{l}{Other features}      						 							& 10 			   		& 0.09547					 	& 0.02087 \\ \hline
	\end{tabular}
\end{table*}

\subsection{Proof of Proposition \ref{prop:etg}}
\label{app:etg}

We first note the following fact, for every $t \geq 0$ and $N(0) = 0$,
$$
\PC{N(t) = 0 }{ \lambda(0)=a} = \exp\left(-\frac{a}{\beta}\left(1-e^{-\beta t}\right)\right).
$$

Let $n$ be a scaling parameter and let $c_n$ be a positive sequence. Then, 
$$
\PC{N\left(\frac{c_n}{\alpha}\right) = 0}{ \lambda(0)=\alpha n} = \exp\left(- n (1-\rho_1) (1-e^{-\frac{c_n}{1-\rho_1}})\right).
$$
Hence, the event $\{N(t)= 0\}$ occurs with exponentially small probability in $n$, when $\lambda(0) = \alpha n$, and $t = c_n/\alpha$, for any $c_n$ such that $c_n = \Omega(1)$.

In particular, we have
$$
\PC{N(\tau_\gamma) = 0}{ \lambda(0)=\alpha n} = \exp\left(-n(1-\rho_1)\left(1-\left(1-\gamma\right)^\frac{1}{1-\rho_1}\right)\right).
$$

Let 
$$
f_\gamma(a) = \EC{T_\gamma}{ \lambda(0) = a}.
$$ 

Then, note

\begin{eqnarray*}
f_\gamma(a) &=& \EC{T_\gamma \mathbf{1}_{\{N(\tau_\gamma) > 0\}}}{ \lambda(0)=a}\\
&& + \EC{T_\gamma \mathbf{1}_{\{N(\tau_\gamma) = 0\}}}{ \lambda(0)=a}\\
&=&  \EC{T_\gamma \mathbf{1}_{\{N(\tau_\gamma) > 0\}}}{ \lambda(0)=a} \\
&& + \left(\tau_\gamma + f_\gamma\left(a e^{-\beta\tau_\gamma}\right)\right)\PC{N(\tau_\gamma)=0}{ \lambda(0)=a}.
\end{eqnarray*}

Now, note
\begin{eqnarray*}
&& \EC{T_\gamma \mathbf{1}_{\{N(\tau_\gamma) > 0\}}}{ \lambda(0) = a}\\ 
&=& \int_0^\infty \PC{T_\gamma \mathbf{1}_{\{N(\tau_\gamma) > 0\}} > t}{ \lambda(0)=a}dt\\
&\leq & \tau_\gamma \PC{N(\tau_\gamma) > 0}{ \lambda(0) = a}\\
&& + \int_{\tau_\gamma}^\infty \PC{T_\gamma \mathbf{1}_{\{N(\tau_\gamma) > 0\}} > t}{ \lambda(0)=a}dt,
\end{eqnarray*}

and, for $t \geq \tau_\gamma$, 
\begin{eqnarray*}
&& \PC{T_\gamma \mathbf{1}_{\{N(\tau_\gamma) > 0\}} > t}{ \lambda(0)=a}\\ 
&= & \PC{0 < N(\tau_\gamma), N(t)  < \gamma N(+\infty)}{ \lambda(0)=a}\\
&\leq & \EC{\frac{\gamma N(+\infty)}{N(t)}\mathbf{1}_{\{N(\tau_\gamma) > 0\}}}{ \lambda(0) = a}\\
&=& \frac{\gamma}{\alpha}\EC{\frac{\lambda(t)}{N(t)} \mathbf{1}_{\{N(\tau_\gamma) > 0\}}}{ \lambda(0) =a}\\
&\leq & \frac{\gamma}{\alpha} \EC{\frac{\lambda(\tau_\gamma)}{N(\tau_\gamma)}e^{-\alpha(t-\tau_\gamma)}\mathbf{1}_{\{N(\tau_\gamma) > 0\}}}{ \lambda(0) = a}.
\end{eqnarray*}

It follows that
\begin{eqnarray*}
&& \EC{T_\gamma \mathbf{1}_{\{N(\tau_\gamma) > 0\}}}{ \lambda(0) = a}\\ & \leq &  \tau_\gamma \PC{N(\tau_\gamma) > 0 }{ \lambda(0) = a}\\
&& + \gamma \frac{1}{\alpha}\EC{\frac{\lambda(\tau_\gamma)}{\alpha N(\tau_\gamma)}\mathbf{1}_{\{N(\tau_\gamma) > 0\}}}{ \lambda(0) = a}.
\end{eqnarray*}

Putting the pieces together, we have
\begin{eqnarray*}
f_\gamma(a) & \leq & \frac{1}{\alpha}\left(\log\left(\frac{1}{1-\gamma}\right) + \gamma \EC{\frac{\lambda(\tau_\gamma)}{\alpha N(\tau_\gamma)} \mathbf{1}_{\{N(\tau_\gamma) > 0\}} }{ \lambda(0)=a}\right)\\
&& + f_\gamma\left(a\left(1-\gamma\right)^{\frac{1}{1-\rho_1}}\right)\PC{N(\tau_\gamma) = 0 }{ \lambda(0) = a}.
\end{eqnarray*}

\subsection{Proof of Corollary~\ref{prop:etg1}}
\label{app:etg1}

First, note
\begin{eqnarray*}
\EC{\frac{\lambda(\tau_\gamma)}{\alpha N(\tau_\gamma)} \mathbf{1}_{\{N(\tau_\gamma) > 0\}} }{ \lambda(0)=a}
&\leq & \frac{1}{\alpha}\EC{\lambda(\tau_\gamma)}{ \lambda(0)=a}\\
&=& \frac{1}{\alpha} a e^{-\alpha \tau_\gamma}\\
&=& \frac{1}{\alpha} a (1-\gamma)\\
&=& \frac{a}{\alpha n}\\
&= & 1.
\end{eqnarray*}

Second, note that $a(1-\gamma)^{\frac{1}{1-\rho_1}} = \alpha n^{-\frac{\rho_1}{1-\rho_1}} = o(1)$. Combining this with $f_\gamma(0) = 0$ for all $\gamma \in [0,1]$, we have $f_{1-1/n}\left(\alpha n^{-\frac{\rho_1}{1-\rho_1}}\right) = o(1)$.

The assertion of the corollary follows by combining the above observations with the bound in Proposition~\ref{prop:etg}.

\subsection{Importance of predictive features}
\label{app:features}
All 1889 features we used in our experiments could be categorized in the following groups:

\textbf{Content features} are static properties of the post, such as the type of media it contains, language of the text, and number of mentioned users.\

\textbf{Page features} are properties of the page that posted the content, such as the number of followers, fans, and number of posts published last month.

\textbf{Engagement features} describe the cumulative history of users' interactions with the post, such as comments, shares, reactions and views. We count them using different time windows and starting points, e.g., number of comments in the last hour, number of shares during the first day since it was published, number of views per minute in the last 15~minutes, etc. The number of features in this category is large since we used a cross product of all possible engagement types, time window sizes and starting points, etc. We also added a few combination features here, which are the ratios of counters for different types of engagement (i.e., comments to shares). Another subgroup in this category consists of cumulative view count features on the page's previous posts, taken at different points in time before prediction.

\textbf{Other features} category contains a handful of features which did not belong to any of the above-mentioned categories, including prediction time, content age at the time of prediction, number of group members if the post was published in a group, etc.

Table~\ref{app:tab:features} shows the cardinality of each feature category as well as its cumulative importances (permutation importances over the test set) for both of the models. For the model predicting cascade size at reference horizon $\delta^*$, cumulative importance of the most important subgroup --- views on the post --- is around 53\%. For the model predicting effective growth exponent $\alpha$, two most important feature subgroups --- page features and the page-level engagement features --- have cumulative importance of 32\% each.

\subsection{Tuning reference horizon parameter $\delta^*$}
\label{sec:tuning}

\begin{figure}
	\begin{center}
		\includegraphics[width=1.0\linewidth]{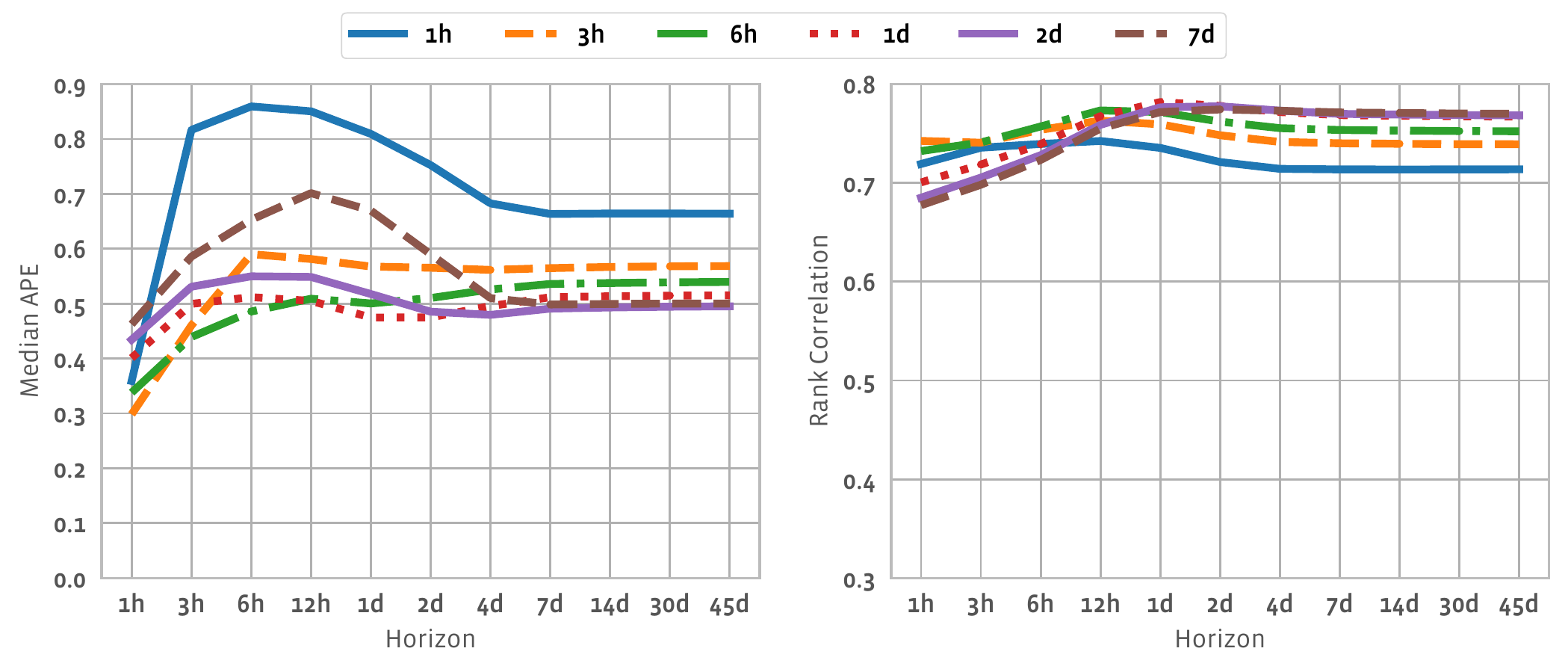}
		\caption{Sensitivity of the model to the choice of $\delta^*$.}
		\label{fig:prediction_deltacap}
	\end{center}
\end{figure}

The key hyper-parameter we need to choose for our model is the reference horizon $\delta^*$ which is used for the point predictor $Y(\delta^*;s)$ (Eq.~(\ref{equ:ystar})). From Figure~\ref{fig:prediction_deltacap}, we observe that the models with very small  $\delta^*$, i.e., $1$h and $3$h, perform poorly on both metrics for long horizons. However, the gains in performance become less significant when $\delta^*$ increases over $24$h. The opposite is true for short horizons: The best performing models in the initial hours after predictions are the $3$h and $6$h models. Evidently, a choice of $\delta^*$ allows us to trade-off between the performance on short and long horizons. We choose the best performing models with a single (HWK (1d)), double (HWK (6h,4d)) and triple (HWK (6h,1d,4d)) point estimators for the experiments in this section by minimizing the Median APE across all horizons.

\subsection{Conditioning on the content popularity} 
\label{sec:performance_cascade_size}

\begin{figure}
	\begin{center}
		\includegraphics[width=\linewidth]{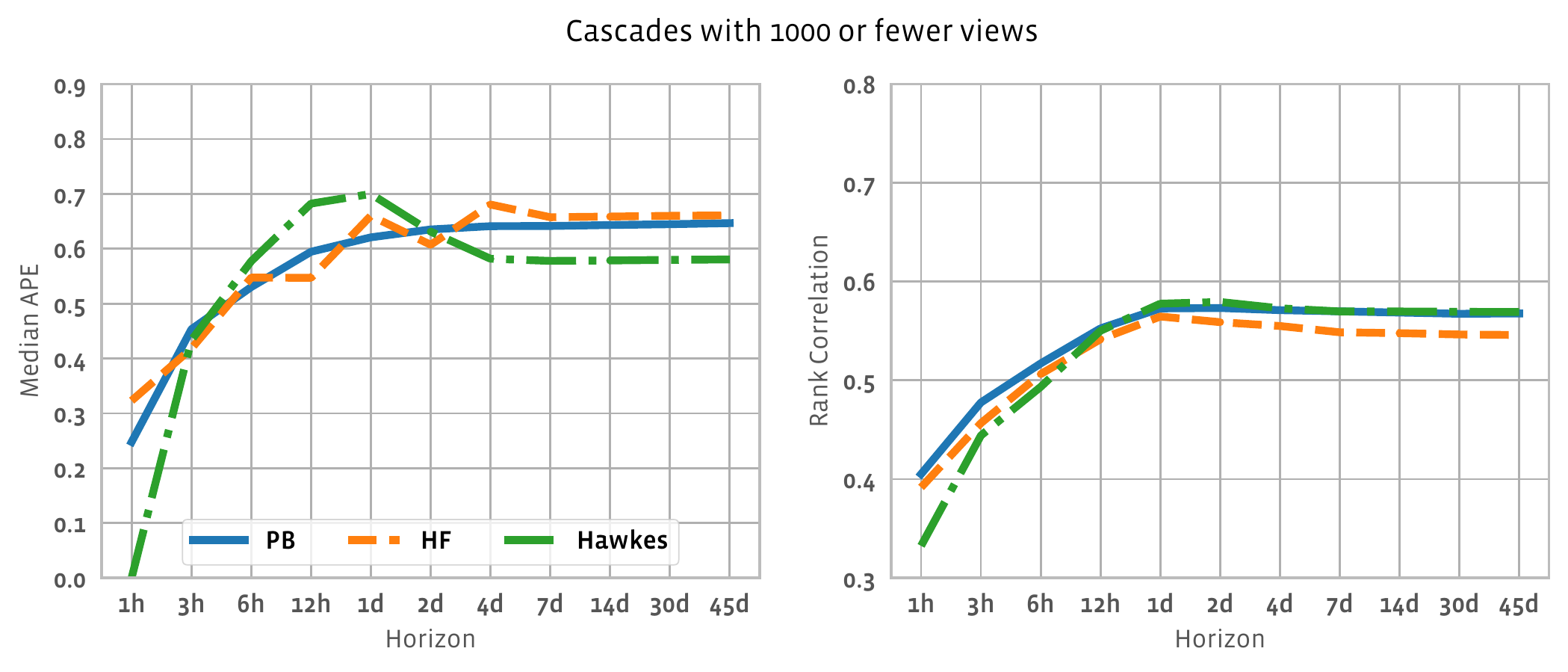}
		\includegraphics[width=\linewidth]{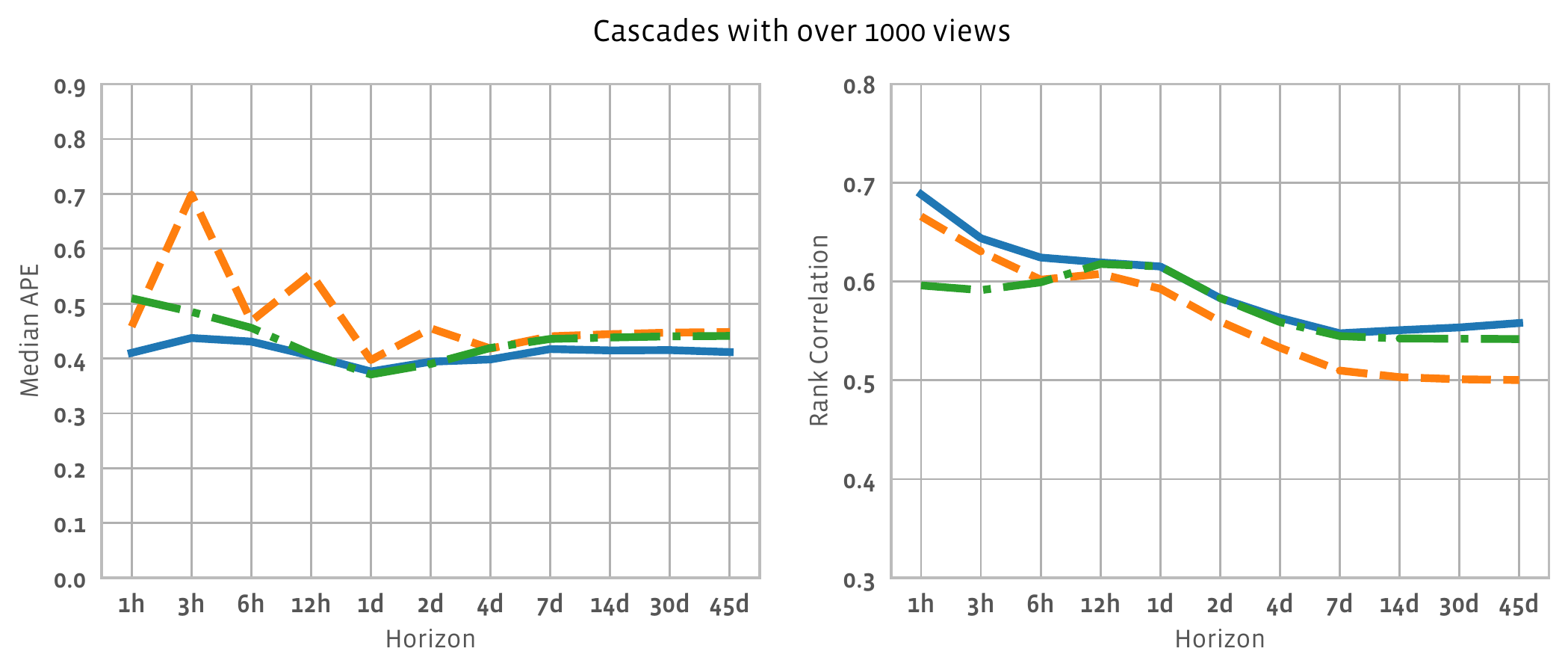}
		\caption{Performance for small and large \dkpolish{cascades}.}
		\label{fig:prediction_size}
	\end{center}
	
\end{figure}

We further examine the relative performance of our model conditioned on the true popularity of the content item. We notice that the performance gain of our model on long horizons (here we consider the best performing model variant HWK (6h,1d,4d)) is particularly evident for small cascades (top-left plot in Figure~\ref{fig:prediction_size}). However, the largest percentage errors on medium horizons (i.e., between $6$h and $1$d) are also mainly featured in the small cascades. Intuitively, the same absolute error corresponds to a large percentage error on a smaller cascade than on a larger one. This is supported by the observation, that all of the considered methods feature significantly better Median APE performance on the larger cascades (bottom-left plot) than on the smaller ones (top-left plot). 

\end{document}